\documentclass[11pt,a4paper,reqno,intlimits,sumlimits]{amsart}

\usepackage{amssymb}
\usepackage{chicago}
\usepackage{color}
\usepackage[mathscr]{euscript}
\usepackage{xspace}
\usepackage{enumerate}
\usepackage{epsfig,rotating,booktabs}

\graphicspath{{./graphs/}}
\newtheorem{theorem}{Theorem}[section]    % the numbering will be chapternr.nr; if you put [section], then the numbering becomes chapternr.sectionnr.nr
\newtheorem{lemma}[theorem]{Lemma}
\newtheorem{proposition}[theorem]{Proposition}

\newtheorem{corollary}[theorem]{Corollary}

\theoremstyle{definition}
\newtheorem{remark}[theorem]{Remark}

\newtheorem{definition}[theorem]{Definition}
\newtheorem{example}[theorem]{Example}

\renewenvironment{proof}{{\parindent 0pt \it{ Proof:}}}{\mbox{}\hfill\mbox{$\Box\hspace{-0.5mm}$}\vskip 16pt}

\newenvironment{proofprop}[1]{{\parindent 0pt \it Proof of Proposition #1:}}{\mbox{}\hfill\mbox{$\Box\hspace{-0.5mm}$}\vskip 16pt}

\def\lev{L\'{e}vy\xspace}

\newcommand{\cB}{{\mathcal{B}}}
\newcommand{\cN}{{\mathcal{N}}}
\newcommand{\cZ}{{\mathcal{Z}}}

\newcommand{\cG}{{\mathcal{G}}}

\newcommand{\cI}{{\mathcal{I}}}
\newcommand{\cT}{{\mathcal{T}}}

\newcommand{\cO}{{\mathcal{O}}}
\newcommand{\cP}{{\mathcal{P}}}

\newcommand{\bG}{{\mathbb{G}}}

\newcommand{\er}{{\mathbb{R}}}
\newcommand{\R}{\er}
\newcommand{\erd}{{\mathbb{R}^{d}}}

\newcommand{\en}{{\mathbb{N}}}

\newcommand{\ii}{\operatorname{i}\kern -0.8pt}

\newcommand{\E}{{\mathbb{E}}}
\newcommand{\Q}{{\mathbb{Q}}}
\renewcommand{\P}{{\mathbb{P}}}

\def\ud{\ensuremath{\mathrm{d}}}
\def\dt{\ud t}
\def\ds{\ud s}
\def\dx{\ud x}
\def\dy{\ud y}

\def\dz{\ud z}

\newcommand{\la}{\langle}
\newcommand{\ra}{\rangle}

\newcommand{\set}[1]{\ensuremath{\left\{#1\right\}}}
\newcommand{\sset}[1]{\ensuremath{\{#1\}}}
\newcommand{\indik}{{\mathbf{1}}}
\newcommand{\ind}[1]{\indik_{\sset{#1}}}

\newcommand{\ifL}[2]{\ensuremath{\mathbf{1}_{\{L_{#1} \leq #2\}}}}

\begin{document}

\title[Credit portfolio modeling]{Discrete tenor models for credit risky portfolios driven by time-inhomogeneous L\'evy processes}

\author{Ernst Eberlein}
\author{Zorana Grbac}
\author{Thorsten Schmidt}
\address{Ernst Eberlein, University of Freiburg, Department of Mathematical Stochastics, Eckerstr. 1, 79104 Freiburg, Germany}
\email{eberlein@stochastik.uni-freiburg.de}
\address{Zorana Grbac, Institute of Mathematics, TU Berlin, Str. des 17. Juni 136, 10623 Berlin, Germany}
\email{grbac@math.tu-berlin.de}
\address{Thorsten Schmidt, Chemnitz University of Technology, Reichenhainer Str. 41, 09126 Chemnitz, Germany}
\email{thorsten.schmidt@mathematik.tu-chemnitz.de}

\thanks{The research of Z.G. benefited from the support of the \emph{Chaire Risque de Cr\'edit}, F\'ed\'eration Bancaire Fran\c{c}aise and the DFG Research Center MATHEON. We would like to thank the associate editor and two anonymous referees for their valuable remarks.}
%\subjclass[2000]{}

\date{\today}\maketitle\pagestyle{myheadings}\frenchspacing

\begin{abstract}
The goal of this paper is to specify dynamic term structure models with discrete tenor structure for credit portfolios in a top-down setting driven by time-inhomogeneous L\'evy processes. We provide a new framework, conditions for absence of arbitrage, explicit examples, an affine setup which includes contagion and pricing formulas for STCDOs and options on STCDOs. A calibration to iTraxx data with an extended Kalman filter shows
an excellent fit over the full observation period. The calibration is done on a set of CDO tranche spreads ranging across six tranches and three maturities.
\end{abstract}

\section{Introduction}

Contrary to the single-obligor credit risk models, portfolio credit risk models consider a pool of credits consisting of  different obligors  and the adequate  quantification of risk for the whole portfolio becomes a challenge. A good model for portfolio credit risk should incorporate two components:  default risk, which includes in particular the dependence structure in the portfolio (also termed  default correlation), and spread risk, which represents the risk related to changes of interest rates and  changes in the credit quality of the obligors.

The main application of such a portfolio model  which we discuss in Section 8 is the valuation of tranches of \emph{collateralized debt obligations} (CDOs) and related derivatives. We would like to emphasize that variants of this model can be used for the valuation of other asset-backed securities.
Currently, due to the sovereign credit crisis that has affected Europe, the issuance of so-called European Safe Bonds (ESBs) is discussed, where the underlying portfolio would consist of  sovereign bonds of EU member states with fixed weights set by a strict rule which is proportional to GDP. Our model is easily adapted for pricing of such and other similar asset-backed securities whatever the precise
specification of these instruments would be.

Generally speaking, CDOs are structured asset-backed securities, whose value and payments depend on a pool of
underlying assets - such as bonds or loans - called the \emph{collateral}. They consist of different \emph{tranches} representing different risk classes, ranging from \emph{senior} tranches with the lowest risk, over \emph{mezzanine} tranches, to the \emph{equity} tranche which carries the highest risk. If
defaults occur in the collateral, the corresponding losses are transferred to investors  in order of seniority, starting with the equity tranche.

Among various portfolio credit risk models, there are two main approaches to be distinguished: the \emph{bottom-up} approach where the default event of each individual obligor is modeled, and the \emph{top-down} approach where the aggregate loss process of a given portfolio is modeled and the individual obligors in the portfolio are not identified.
For a detailed overview of bottom-up and top-down approaches we refer to \citeN{LiptonRennie11} and \citeN{BieleckiCrepeyJeanblanc10}.
The latter approach was investigated in a series of recent papers, among which we mention \citeN{Schoenbucher05}, \citeN{SideniusPiterbargAndersen08}, \citeANP{EhlersSchoenbucher06} (\citeyearNP{EhlersSchoenbucher06}, \citeyearNP{EhlersSchoenbucher09}), \citeN{ArnsdorfHalperin08}, \citeN{LongstaffRajan08}, \citeN{ErraisGieseckeGoldberg10}, \citeN{FilipovicOverbeckSchmidt11} and \citeN{ContMinca12}.

{In this paper we present a dynamic term structure model with \emph{discrete tenor structure} which studies portfolio credit risk in a top-down setting. The framework is developed in the spirit of the so-called \emph{Libor market model}. The need for such an approach is illustrated in \citeN{Carpentier09}, and to our knowledge only \citeN{BennaniDahan04} studied such models for CDOs.
As in \citeN{FilipovicOverbeckSchmidt11} we  utilize  $(T, x)$-bonds. In that paper a dynamic Heath-Jarrow-Morton (HJM) forward spread model for $(T,x)$-bonds has been analyzed under the
assumption that $(T,x)$-bonds are traded for all maturities $T \in [0, T^*]$. Here we acknowledge the fact, that the set of traded maturities is only finite. This has important consequences for modeling and we introduce a new framework which takes this fact into account. We show that this framework possesses some clear advantages.}

The first major difference is due to the fact that in the no-arbitrage condition in Theorem \ref{th:drift-cond} one has to
consider only finitely many maturities $T_k$. The HJM-approach instead has to guarantee the validity
of this condition for a continuum of maturities. This restricts the model in an unnecessary way since
traded products are only available for a small number of maturities. As we will show in the examples in section \ref{sec:examples}
one gains considerable additional freedom in the specification of arbitrage-free models. See in particular Remark
\ref{rem:affineFOS}. The second difference is that we are able to include a contagion effect in an
affine specification of this approach. It is evident that contagion is an important issue in the current crises.
It should be mentioned that a model with only finitely many maturities can be extracted from the
HJM framework, see \citeN{SchmidtZabczyk12}, which of course inherits the HJM-properties.

As driving processes for the dynamics of credit spreads, a wide class of time-inhomogeneous L\'evy processes is used. This allows for jumps in the spread dynamics which are not only triggered by  defaults in the underlying portfolio. In fact the empirical study in \citeN{ContKan11} reveals that jumps in the spread dynamics do not only occur at the default dates of the obligors in the portfolio, but they can also be caused by a macroeconomic event which is external to the portfolio.
In \citeN{ContKan11} the  bankruptcy of Lehman Brothers is given as an example of such an event. This is a weak point of some of the recently proposed portfolio credit risk models in which jumps in the spread dynamics occur only at default dates in the underlying portfolio (see a detailed discussion in \citeN{ContKan11}).
In the model  developed in the sequel we incorporate both types of jumps in the spread dynamics.

The model is calibrated to iTraxx data from January 2008 to August 2010 applying an
extended Kalman filter to a two-factor  affine diffusion specification of our approach, as proposed in \citeN{EksiFilipovic12}. Contrary to the usual calibration to data from one day (see \citeN{ContDeguestKan10} for an overview and excellent empirical comparison), we calibrate the model to a much larger dataset running over three years. Already in the simple two-factor diffusion case a very good performance across different tranches
and maturities is achieved.

The paper is structured as follows. In Section 2 we introduce the setting and basic notions. In Section 3 we describe the aggregate CDO loss process $L$ and the driving process $X$ and specify the dynamics of the credit spreads.
Section 4 reviews the forward martingale measure approach. Section 5 contains the main results on the absence of arbitrage and Section 6 examines these results in a series of explicit examples. In Section 7 we focus  attention on an affine specification which is able to incorporate contagion effects. In Section 8 we show how the valuation of derivatives can be facilitated by introducing appropriate defaultable forward measures and present a valuation formula for a single tranche CDO, which is the standard instrument for investing in a CDO. Moreover, we study the valuation of call options on STCDOs.
Finally, in Section 9 we propose a two-factor affine specification and calibrate it to data from the iTraxx series.

\section{Basic notions and definitions}
\label{sec:CDO-2}

Let $T^{*} >0$ be a fixed time horizon and let a complete stochastic basis $(\Omega, \cG, \bG, \Q_{T^*})$ be given, where $\cG=\cG_{T^*}$ and $\bG=(\cG_{t})_{0\leq t \leq T^*}$ is some filtration satisfying the usual conditions. For simplicity we write $\Q^*$ for $\Q_{T^*}$. The expectation with respect to \ $\Q^*$ is denoted by $\E^*$. The filtration $\bG$ represents the filtration which contains all the information available in the market. All the price and interest rate processes in the sequel are adapted to it.
Furthermore, assume that the tenor structure  $0=T_{0} < T_{1} < \ldots < T_{n}=T^{*}$ is given. Set  $\delta_{k} := T_{k+1} - T_{k}$, for $k=0, \ldots, n-1$.

We assume that default-free zero coupon bonds with
maturities $T_{1}, \ldots, T_{n}$ are traded in the market and denote by $P(t, T_{k})$ the time-$t$ price of a
default-free zero coupon bond with maturity $T_{k}$. For default-free zero coupon
bonds $P(T_{k}, T_{k})=1$ for all $k$. Furthermore we assume that $P(t,T_k)>0$ for any $0 \le t \le T_k$ and all $k$.

Furthermore, there is  a \emph{pool} of credit risky assets and we denote by $L=(L_t)_{t \ge 0}$ the nondecreasing \emph{aggregate} \emph{loss process}.  Assume that the total nominal is normalized to $1$ and
 denote by $\cI:=[0, 1]$ the set of loss fractions such that $L$ takes values in $\cI$.

\begin{remark}
\label{rem:bottom-up}
This approach is called \emph{top-down} as we model the aggregate loss process directly. In the \emph{bottom-up} approach
one models instead the individual default times: for this, denote by $\tau_1,\dots,\tau_m$ the default times
of the credit risky securities in the collateral and their (possibly random) loss given default by $q_1,\dots,q_m$. Then
$$
L_t = \sum_{i =1}^m q_i \indik_{\set{\tau_i \leq t}}.
$$
\end{remark}
\begin{remark}
The filtration $\bG$ denotes the full market filtration to which the aggregate loss process is adapted. In \citeN{EhlersSchoenbucher09} the full market filtration is constructed as a progressive enlargement of a default-free filtration (known as a background or a reference filtration) with the default times in the portfolio
 under a certain version of the immersion hypothesis. Note that here $\bG$ is general and we do not restrict ourselves to the case studied in \citeN{EhlersSchoenbucher09}. In particular, the immersion hypothesis is not needed.
\end{remark}

\begin{definition}
A security which pays  $\ifL{T_k}{x}$ at  $T_{k}$ is called  $(T_{k}, x)$-\emph{bond}.
Its price at time $t \le T_k$ is denoted by $P(t,T_k,x)$. Note that $P(t,T_k,x) = 0$ on the set $\{L_{t} >x\}$.
\end{definition}
If the market is free of arbitrage, $P(t, T_{k}, x)$ is nondecreasing in $x$ and
\begin{equation}
\label{eq:risk-free-bond-1}
P(t, T_{k},1) = P(t, T_{k}).
\end{equation}
In \citeN{FilipovicOverbeckSchmidt11} a forward rate model for $(T,x)$-bonds has been analyzed under the
assumption that $(T,x)$-bonds are traded for all maturities $T \in [0, T^*]$. Here we acknowledge the fact, that in practice the set of maturities for which the bonds are traded is finite.

\begin{definition}
The $(T_k,x)$-forward price is given by
\begin{align}\label{eq:forwardpriceprocess}
F(t,T_k,x) :=  \frac{P(t,T_k,x)}{P(t,T_{k})}
\end{align}
for $0 \le t \le T_k$.
\end{definition}

The   $(T_k,x$)-forward prices actually give the
distribution of $L_{T_k}$ under the $\Q_{T_k}$-forward measure which will be defined later in   \eqref{eq:measure-change-default-free-CDO}.
Indeed, note that if we take $P(\cdot,T_k)$ as the numeraire we obtain
\begin{align*}
\Q_{T_k}\big( L_{T_k} \le x |\cG_t \big) &= \frac{1}{P(t,T_k)} \, P(t,T_k) \E_{\Q_{T_k}}\big( \ind{L_{T_k} \le x} |\cG_t \big) \\
&= \frac{P(t,T_k,x)}{P(t,T_k)} = F(t,T_k,x).
\end{align*}
Furthermore, we set for $k \in \{0,\dots,n-1\}$ and $t \le T_k$, on $\{L_t \le x\}$,
\begin{align}
\label{eq:connection-H-F}
 H(t,T_k,x)&  := \frac{F(t,T_{k+1},x)}{F(t,T_k,x)}.
 \end{align}
This quantity relates to credit spreads as follows:
intuitively, the credit spread quantifies the additional yield above the risk-free rate which the holder of a $(T_k,x)$-bond receives in compensation for
taking the risk that $L$ jumps over the level $x$.
Recall that for the classical Libor rate, with $\delta_k = T_{k+1}-T_k$,
$$ 1+  \delta_k \cdot LIBOR (t,T_k) = \frac{P(t,T_k)}{P(t,T_{k+1})}. $$
If the credit spread is denoted by $cs(t,T_k,x)$, then on $\{L_t \le x\}$
\begin{align}\label{eq:creditspread}
\big( 1 + \delta_k cs(t,T_k,x) \big) \, \big( 1+ \delta_k LIBOR(t,T_k) \big) = \frac{P(t,T_k,x)}{P(t,T_{k+1},x)},
\end{align}
and
\begin{align*}
H(t,T_k,x)^{-1} =  1 + \delta_k cs(t,T_k,x) = \frac{P(t,T_k,x)}{P(t,T_{k+1},x)} \frac{ P(t,T_{k+1})}{P(t,T_k)}.
\end{align*}
 As we shall see in Section \ref{sec:pricing}, the quantities $H(t,T_k,x)$
 and not the credit spreads $cs(t,T_k,x)$ appear as the main ingredients in pricing formulas for portfolio credit derivatives.

 By induction we obtain the following decomposition of the $(T_k, x)$-forward price.
 For $t\in [0, T^*]$, let $j(t):= \inf\{i \in \en:  T_{i-1}<t\le T_i\}$, with the convention $j(0)=0$,
 denote the
 unique integer $j$ such that $T_{j-1}< t \le T_j$.
 From \eqref{eq:connection-H-F} we obtain
 \begin{align}\label{def:F}
 F(t,T_k,x) & = \ind{L_t \le x} F(t,T_{j(t)},x) \prod_{i=j(t)}^{k-1} H(t,T_i,x).
\end{align}

Summarizing, the model has   three ingredients to be specified: the dynamics of the loss process $L$, the credit spread via  $H$ and
the $F(t,T_{j(t)},x)$. This of course should be done in a way which excludes arbitrage
and leads to tractable pricing formulas. Both points will be discussed in the next sections.

%%%%%%%%%%%%%%%%%%%%%%%%%%%%%%%%%%%%%%%%%%%%%%%%%%%%%%%%%%%%%%%%%%%%%%%%%%%%%%%%%%%%%%
%
% The driving process
%
%%%%%%%%%%%%%%%%%%%%%%%%%%%%%%%%%%%%%%%%%%%%%%%%%%%%%%%%%%%%%%%%%%%%%%%%%%%%%%%%%%%%%%

\section{Ingredients of the model}
\label{sec:driving-process}

Let us now describe the processes which drive the model. A realistic assumption is that the dynamics of defaultable quantities related to the assets in the given portfolio is influenced by the aggregate loss process $L$. This means that when a default occurs in the portfolio, the default intensities of the other assets may be affected as well.
In order to incorporate these features, we design a model where two sources of randomness appear:
\begin{itemize}
\item[(1)] a time-inhomogeneous \lev process $X$ representing the market randomness, which is driving the default-free and the pre-default dynamics
\item[(2)] the aggregate loss process $L$ for the given pool of credits.
\end{itemize}
From now on we assume that these two processes are independent with c\`adl\`ag trajectories. Note that this implies that there are no simultaneous jumps of $X$ and $L$. The independence assumption can be relaxed at the cost of having less explicit expressions.
However, joint jumps in credit spreads and the loss process are incorporated
via an explicit contagion mechanism, see \eqref{def:H}.

 The definition and main properties of time-inhomogeneous \lev processes can be found for example in \citeN{EberleinKluge06a}. We recall that these processes are also known as
 processes with independent increments and absolutely continuous characteristics
 (PIIAC, cf. \citeN{JacodShiryaev03}), or additive processes in the sense of \citeN{Sato99}. For general semimartingale theory we refer to the book by \citeN{JacodShiryaev03}, whose notation we adopt throughout the paper. Time-inhomogeneous \lev processes have already been used in term structure modeling of interest rates because of their analytical tractability combined with a high degree of flexibility, which allows for an adequate fit of
the term structure of volatility smiles, i.e. of the change of the smile across maturities; see \citeN{EberleinKluge06a}  and \citeN{EberleinKoval06}. In credit risk modeling there is also evidence that processes with jumps are a convenient choice as driving processes for the dynamics of credit spreads; see \citeN[p. 118]{ContKan11}, where the observation that the jumps in
the spreads are not only tied to defaults in the underlying portfolio is stated.

Before giving a precise characterization of the driving process, let us describe the aggregate loss process $L$ in more detail. We assume that $L_{t}=\sum_{s \leq t} \Delta L_{s}$ is an $\cI$-valued nondecreasing marked point process with absolutely continuous $\Q^*$-compensator
\begin{equation}
\label{eq:comp-A}
\nu^{L}(\dt, \dy) = F^{L}_{t}(\dy) \dt,
\end{equation}
where  $F^{L}$ is a transition kernel from $(\Omega \times [0, T^*], \cP)$ into $(\er, \cB(\er))$ and $\cP$ denotes the predictable
$\sigma$-algebra on $\Omega \times [0, T^*]$.

Note that $L$ is a semimartingale with finite variation and with canonical representation
$$
L= x \ast \mu^{L} = x \ast (\mu^{L}- \nu^{L}) +  x \ast \nu^{L},
$$
where $\mu^{L}$ denotes its random measure of jumps. Moreover,  $L$ is a special semimartingale since its jumps are bounded by 1.

The indicator process $\ifL{t}{x}$ is a c\`{a}dl\`{a}g, decreasing process with intensity process
\begin{equation}\label{eqn:lambdatx}
\lambda(t, x) = F_t^{L}((x-L_{t}, 1] \cap \cI);
\end{equation}
i.e.\ the process
\begin{equation}
\label{eq:indik_martingale}
M_{t}^{x}= \ifL{t}{x} + \int_{0}^{t} \ifL{s}{x} \lambda(s, x) \ds
\end{equation}
is a $\Q^*$-martingale (see \citeN{FilipovicOverbeckSchmidt11}, Lemma 3.1).

Let us provide an example for the loss process $L$. Note that the process defined in Remark \ref{rem:bottom-up} is also an example for $L$.
\begin{example}
 Consider a compound Poisson process $Z=(Z_t)_{t\geq 0}$ with only positive jumps, defined as follows
 % (cf. Theorem 4.3 in \citeN{Sato99}):
$$
Z_t = \sum_{i=1}^{N_t} Y_i, \qquad Z_0=0,
$$
where $N=(N_t)_{t \geq 0}$ is a Poisson process with intensity $c$, and $Y_i$, $i\in \en$, are mutually independent, identically distributed random variables, independent of $N$, with distribution $P^Y$ on $\er^+$ (e.g. take $P^Y$ to be a Gamma or an exponential distribution). The L\'evy measure of $Z$ is given by $F^Z= c P^Y$.
Next, we define the process $L=(L_t)_{t\geq 0}$ by
$$
L_t:= f(Z_t),
$$
where $f:\er^+ \to [0, 1]$ is given by $f(x) = 1- e^{-x}$. Since $f$ is a nondecreasing function, $L$ is a nondecreasing process taking values in $[0,1]$. Moreover, it is a pure-jump process by definition. The jumps of $L$ are given by
$$
\Delta L_t = e^{-Z_{t-}} f(\Delta Z_t).
$$
Hence, $F^L_t$ equals
\begin{equation}
\label{eq:F-example}
F^{L}_t (E)  = \int_{\er^+} \indik_E (e^{-Z_{t-}} f(x)) F^Z(\dx) = \int_{\er^+} c \indik_E (e^{-Z_{t-}} f(x)) P^Y(\dx),
\end{equation}
for $E \in \cB(\er^+ \setminus \set{0})$, which completes the example.
\end{example}

Let $X$
be an $\er^{d}$-valued
time-inhomogeneous \lev process on the stochastic basis $(\Omega, \cG, \bG, \Q^*)$ with $X_0=0$ a.s. and canonical representation given by
\begin{equation}
\label{eq:driving_process}
X_{t} =W_{t} + \int_{0}^{t} \int_{\erd} x (\mu - \nu) (\ds, \dx);
\end{equation}
where $W$ is a $d$-dimensional standard Brownian motion with respect to $\Q^*$, $\mu$ is the random measure of jumps of $X$ and $\nu$ such that $\nu(\dt, \dx)=F_t(\dx)\dt$ is its $\Q^*$-compensator. To ensure the existence of representation \eqref{eq:driving_process} we assume
\begin{itemize}
\item[$(\mathbf{A1})$]
There exist constants $\tilde{C}, \varepsilon > 0$ such that
$$
\sup_{0\leq t\leq T^{*}} \left( \, \int_{|y| > 1} \exp \langle u, y \rangle F_{t}(\ud y) \right) < \infty,
$$
for every $u \in [-(1+\varepsilon)\tilde{C}, (1+\varepsilon)\tilde{C}]^{d}$.
\end{itemize}
This assumption entails the \emph{existence of exponential moments} of $X$, i.e. $\E^*[\exp \langle u, X_{t} \rangle] < \infty$, for all $t \in [0, T^*]$ and $u$ as above; cf. Lemma 6 in \citeN{EberleinKluge06a}.

The main ingredient for our model is the specification of the dynamics of the \emph{credit spreads}
via specification of $H$. We assume that
\begin{align} \label{def:H}
 H(t,T_k,x) &= H(0,T_k,x) \exp\bigg( \int_0^t a(s,T_k,x) ds + \int_0^t b(s,T_k,x) dX_s  \nonumber \\
 &+ \int_0^t \int_{\cI} c(s,T_k,x;y) \mu^L(ds,dy) \bigg),
\end{align}
where we impose the following assumptions ($\cO$ and $\cP$ denote respectively the optional and the predictable $\sigma$-algebra on $(\Omega \times [0, T^*])$):
\begin{itemize}
\item[$(\mathbf{A2})$] For all $T_{k}$ there is an $\er_{+}^{d}$-valued process $b(s, T_{k},x)$, which as a function of $(s,x) \mapsto b(s, T_k, x)$ is $\cP \otimes \cB(\cI)$-measurable. Moreover,
$$
\sup_{s\in[0,T^*],x\in\cI,\omega\in\Omega}\sum_{k=1}^{n-1} b^{j}(s, T_{k},x) \leq \tilde{C}
$$
for  every coordinate $j \in \set{1, \ldots, d}$, where $\tilde{C}>0$ is the constant from $(\mathbf{A1})$.
If $s >T_{k}$, then $b(s, T_{k}, x)=0$. \\[-2mm]
\item[$(\mathbf{A3})$] For all $T_k$ there is an $\er$-valued process $c(s, T_{k}, x; y)$, which is called the \emph{contagion} parameter and which as a function of $(s,x,y) \mapsto c(s,T_k,x;y)$ is $\cP \otimes \cB(\cI) \otimes \cB(\cI)$-measurable. We also assume
$$
\sup_{s \leq T_{k}, x, y \in \cI, \omega \in \Omega} |c(s, T_{k}, x; y)| < \infty
$$
and $c(s, T_k,x;y) =0$ for $s> T_k$.
\item[$(\mathbf{A4})$] The initial term structure $P(0, T_{k},x)$ is strictly positive, strictly decreasing in $k$ and satisfies
$$
F(0,T_k,x)=\frac{P(0, T_{k}, x)}{P(0, T_{k})} \geq \frac{P(0, T_{k+1},x)}{P(0, T_{k+1})}=F(0,T_{k+1},x).
$$
\end{itemize}
The drift term $a(\cdot, T_{k}, \cdot)$, for every $T_k$, is an $\er$-valued, $\cO \otimes \cB(\cI)$-measurable process  such that $a(s, T_k, x) =0$, for $s > T_k$, which will be specified later. Note that this together with assumptions $(\mathbf{A2})$ and $(\mathbf{A3})$ implies that $H(t,T_i,x)$ remains constant after $T_i$, i.e. $H(t,T_i,x)= H(T_i,T_i,x)$, for $t\geq T_i $.
\begin{remark}
Specifying the dynamics of $H$ in this way, we allow for two kinds of jumps: the jumps caused by market forces, represented by the time-inhomogeneous \lev process $X$, and the jumps caused  by defaults in the portfolio, represented through the aggregate loss process $L$, which allows for contagion effects.
\end{remark}

\section{The forward measures}
In a short excursion we recall the most important results from default-free Libor models and introduce the %important
forward martingale measures.

 In default-free discrete tenor models the forward martingale measures are constructed by backward induction, together with the forward Libor rates. The measure $\Q^*=\Q_{T^*}=\Q_{T_n}$ plays the role of the forward measure associated with the settlement date $T_{n}$ and is called the \emph{terminal forward measure}. We shall write $W^{T_{n}}$ for $W$ and $\nu^{T_{n}}$ for $\nu$ when we wish to emphasize that $\Q^*$ is the terminal forward measure. %the dependence on the measure $\Q_{T_n}$.

The forward measure $\Q_{T_{k}}$ is defined on $(\Omega, \cG_{T_{k}})$ by its Radon--Nikodym derivative with respect to  $\Q_{T_{n}}$, i.e.
\begin{equation}
\label{eq:measure-change-default-free-CDO}
\frac{\ud \Q_{T_{k}}}{\ud \Q_{T_{n}}} \bigg|_{\cG_t} =  \frac{P(0, T_{n})}{P(0, T_{k})}\frac{P(t,T_k)}{P(t,T_n)}.
\end{equation}
We assume that this density has the following
representation as  a stochastic exponential:
\begin{align}\label{eq:forwarddensity}
\frac{\ud \Q_{T_{k}}}{\ud \Q_{T_{n}}}\bigg|_{\cG_t} = \mathcal{E}_t\bigg( \int_0^{\cdot} \alpha(s,T_k)dW_s +
 \int_0^{\cdot} \int_{\erd}(\beta(s,T_k,y)-1) (\mu-\nu)(ds,dy)\bigg),
\end{align}
where $\alpha \in L(W)$ and $\beta \in G_{\mathrm{loc}}(\mu)$ in the sense of Theorem III.7.23 in \citeN{JacodShiryaev03}; for definitions of $L(W)$ and $G_{\mathrm{loc}}(\mu)$ see the same textbook, page 207 and page 72 respectively.
Then, applying Girsanov's theorem, we deduce that
\begin{equation}
\label{eq:brownian-CDO}
W_{t}^{T_{k}} := W_{t}- \int_{0}^{t}  \alpha(s, T_{k}) \ud s
\end{equation}
is a $d$-dimensional standard Brownian motion with respect to $\Q_{T_{k}}$, and
\begin{equation}
\label{eq:compensator-CDO}
\nu^{T_{k}}(\ud s, \ud y) :=  \beta(s, T_{k}, y) \nu (\ud s, \ud y) = F_s^{T_k}(\dy) \ds,
\end{equation}
is the $\Q_{T_{k}}$-compensator of $\mu$, where $F_s^{T_k}(\dy) = \beta(s, T_{k}, y) F_s(\dy)$.
See \citeN{EberleinOezkan05}, Section 4, pp. 338--342, for the detailed construction of Libor rates which are driven by a \lev process.
%For the classical Libor market model driven by a \lev process
%see \citeN{EberleinOezkan05}, Section 4, pp. 340--341. This paper provides the detailed construction of the L\'evy Libor rates.

We denote by $\nu^{L,T_k}(dt,dx)=F_t^{L,T_k}(dx)dt$ the $\Q_{T_k}$-compensator of the random measure $\mu^L$ of
the jumps of the loss process. The existence of $F_t^{L,T_k}$ follows in the same way as the existence of $F_t^{T_k}$ in \eqref{eq:compensator-CDO}.

\begin{remark}[Constant term structure]
\label{rem:constant-risk-free}
If the price processes for default-free bonds $(P(t,T_k))_{0 \le t \le T_k}$ are constant equal to 1 for every $k=1,\dots,n,$
 all forward measures coincide, i.e.\
\begin{align*}
\Q_{T_1}=\dots=\Q_{T_n}=\Q^*.
\end{align*}\label{rem:constantriskfreeinterest}
\end{remark}

%%%%%%%%%%%%%%%%%%%%%%%%%%%%%%%%%%%%%%%%%%%%%%%%%%%%%%%%%%%%%%%%%%%%%%%%%%%%%%%%%%%%%%
%
% Absence of Arbitrage
%
%%%%%%%%%%%%%%%%%%%%%%%%%%%%%%%%%%%%%%%%%%%%%%%%%%%%%%%%%%%%%%%%%%%%%%%%%%%%%%%%%%%%%%

\section{Absence of arbitrage}
The goal of this section is to identify conditions which guarantee absence of arbitrage in our setting.
It is well-known that the model is free of arbitrage if
all $(T_k,x)$-bonds discounted with a suitable numeraire are local martingales and we choose default-free
bonds as numeraires.

The quantity $F(t,T_{j(t)},x)$ given in \eqref{def:F} is the forward
bond price for the closest maturity from time $t$ (typically less than 3 months). In the following discussion of absence of arbitrage we do not have to consider this particular forward bond price. The reason for this is that the market trades only financial instruments whose first tenor date (payment date) is at least a full tenor period away.
As a consequence, we consider $P(\cdot,T_k,x)$ as traded assets, with $k \in \{2,\dots,n\}$, and  study
the question if $(F(t,T_k,x))_{0 \le t \le T_{k-1}}$ are $\Q_{T_k}$-local martingales for any $k \in \{2,\dots,n\}$.
The following lemma shows that the numeraires can be interchanged arbitrarily.
\begin{lemma}
There is equivalence between:
\begin{itemize}
\item[(a)] For each $k=2, \ldots, n$ the process
$$
(F(t,T_k,x))_{0 \leq t \leq T_{k-1}}
$$
is a $\Q_{T_{k}}$-local martingale.
\item[(b)] For each $k, i=2, \ldots, n$ the process
$$
\Big( \frac{P(t, T_{k}, x)}{P(t, T_{i})} \Big)_{0 \leq t \leq T_{i} \wedge T_{k-1}}
$$
is a $\Q_{T_{i}}$-local martingale.
\end{itemize}
\end{lemma}

\begin{proof}
It suffices to note that for fixed $i, k \in \set{2, \ldots, n}$ such that $i \geq k$ (the other case is treated in the same way) we have
$$
\frac{P(t, T_{k}, x)}{P(t, T_{i})} =F(t,T_k,x)  \frac{P(t, T_{k})}{P(t, T_{i})},
$$
where $F(\cdot,T_k,x)=\frac{P(\cdot,T_k,x)}{P(\cdot,T_k)}$ is a  $\Q_{T_{k}}$-local martingale by (a) and $\frac{P(\cdot, T_{k})}{P(\cdot, T_{i})} $ is the density process of the measure $\Q_{T_{k}}$ relative to $\Q_{T_{i}}$, up to a norming constant (cf. equation \eqref{eq:measure-change-default-free-CDO}). Then $\frac{P(\cdot, T_{k}, x)}{P(\cdot, T_{i})}$ is a $\Q_{T_{i}}$-local martingale by Proposition III.3.8 in \citeN{JacodShiryaev03}. The implication $(a) \Rightarrow (b)$ is thus shown. $(b) \Rightarrow (a)$ is obvious.
\end{proof}

Now regarding the discussion at the beginning of this section, we specify \eqref{def:F} further as follows
\begin{align}\label{def:F1}
F(t,T_k,x) := \ind{L_t \le x} \prod_{i=0}^{k-1} H(t,T_i,x),
\end{align}
for any $0 \le t \le T_{k-1}$ with $H(t, T_i, x)$ given by \eqref{def:H}. Recall that $H(t,T_i,x)$ remains constant for $t > T_i$ by assumption.
We examine conditions for absence of arbitrage, i.e. necessary and sufficient conditions for the $(T_k, x)$-forward price process $F(\cdot, T_{k}, x)$ being a local martingale under the forward measure $\Q_{T_{k}}$,
for $k=2,\dots,n$.

Set
\begin{align}
\label{eq:D}
D(t,T_k,x) &:= \sum_{i=1}^{k-1}a(t,T_i,x)
+ \frac{1}{2} \parallel \sum_{i=1}^{k-1}b(t,T_i,x) \parallel^2  \\
& + \Big\la \sum_{i=1}^{k-1}b(t,T_i,x),  \alpha(t,T_k) \Big\ra \nonumber  \\
& + \int_{\er^d} \Big(e^{\left\la \sum_{ i=1}^{k-1}b(t,T_i,x),  y \right\ra}-1  - \Big\la \sum_{ i=1}^{k-1}b(t,T_i,x),  y \Big\ra \beta(t, T_k, y)^{-1} \Big) F_t^{T_k}(dy) \nonumber,
\end{align}
{where $\alpha$ and $\beta$ were introduced in \eqref{eq:forwarddensity}.
Recall that $\nu^{L, T_k}(\dt,\dx)= F_t^{L,T_k}(dx)dt$ is the $\Q_{T_k}$-compensator of the random measure of jumps $\mu^L$.
Analogously to \eqref{eq:indik_martingale}, we get that
\begin{equation}
\label{eq:intensity-QTk}
M_t^{x, T_k}:= \ifL{t}{x} + \int_{0}^{t} \ifL{s}{x} \lambda^{T_k}(s, x) \ds
\end{equation}
is a $\Q_{T_k}$-martingale, where $\lambda^{T_k}(t,x):= F^{L,T_k}_t((x-L_t,1] \cap \cI)$. By $\lambda^1$ we denote the Lebesgue measure on $\er$. }
\begin{theorem}
\label{th:drift-cond}
Assume that $(\mathbf{A1})$--$(\mathbf{A4})$ are in force and let $k \in \{2, \ldots, n\}$, $x \in \cI$. Then the process
$(F(t,T_k,x))_{0 \le t \le T_{k-1}}$
given by \eqref{def:F1} is a $\Q_{T_{k}}$-local martingale if and only if
\begin{align}
\label{eq:drift_condition}
D(t, T_{k}, x)  & =   \lambda^{T_k}(t, x)
- \int_{\cI} \Big( e^{\sum_{i=1}^{k-1}  c(t,T_i,x;y)}-1 \Big) \indik_{\sset{L_{t-} + y \le x }} F^{L, T_k}_t(dy)
\end{align}
on the set $\set{L_{t} \leq x}$, $\lambda^1 \otimes \Q_{T_k}$-a.s.
\end{theorem}
\begin{remark}\label{rem:affineFOS}
Note that in the HJM term structure models, by considering the continuum of maturities one puts unnecessary restrictions on the model. It is a major advantage of models with  discrete tenor structure  that only those maturities are considered which are traded in the market.
It will become clear in the various examples, which are discussed in Section \ref{sec:examples},
that the drift condition \eqref{eq:drift_condition} can be satisfied while there is still a high
degree of freedom to specify the intensity of the loss process.
This is not the case in the HJM framework, where the risky short rate is directly connected to the intensity of the loss process, see equation (3.11) in \citeN{FilipovicOverbeckSchmidt11}. For example, we are able to specify the dynamics of the spreads and still have an \emph{arbitrary} intensity of the loss process.
Moreover we are able to specify an affine version of the model which includes contagion.
\end{remark}

\begin{proof}
We  calculate first the dynamics of the forward price processes under the forward measures and then derive the drift conditions.
 We fix $x$ and $T_k$ and define
$$ G(t)=G(t,k,x):=\prod_{i=0}^{k-1} H(t,T_i,x) , $$
such that  $F(t, T_{k}, x) = G(t)  \ifL{t}{x}$. Using integration by parts yields
\begin{align*}
\ud F(t, T_{k}, x) & = G(t-)\ud \ifL{t}{x} + \ifL{t-}{x} \ud G(t) + \ud \left[ G,  \ifL{\cdot}{x} \right]_{t}\\
& =:  (1') + (2') + (3').
\end{align*}
We deal separately with each of the above three summands. Regarding $(1')$,
equation \eqref{eq:intensity-QTk} yields
\begin{eqnarray*}
\ud \ifL{t}{x}  & = & \ud M_{t}^{x, T_k} - \ifL{t}{x} \lambda^{T_k}(t, x) \dt\\
& = &  \ifL{t-}{x} \ud M_{t}^{x, T_k} - \ifL{t-}{x} \lambda^{T_k}(t, x) \dt \\
& = & \ifL{t-}{x} \Big(\ud M_{t}^{x, T_k} - \lambda^{T_k}(t, x) \dt\Big),
\end{eqnarray*}
since a short computation shows that $ \ud M_{t}^{x, T_k} = \ifL{t-}{x} \ud M_{t}^{x, T_k}$. Hence,
\begin{eqnarray*}
(1') & = & G(t-)\ifL{t-}{x} \Big(\ud M_{t}^{x, T_k} - \lambda^{T_k}(t, x) \dt\Big) \\
& = &  F(t-, T_{k}, x)  \Big(\ud M_{t}^{x, T_k} - \lambda^{T_k}(t, x) \dt\Big).
\end{eqnarray*}

Regarding (2'), we obtain using  \eqref{def:H}
  \begin{align*}
G(t) &= G(0) \exp\Bigg( \int_0^t \sum_{i=1}^{k-1}a(s,T_i,x) ds \\
& + \int_0^t \sum_{i=1}^{k-1}b(s,T_i,x) dX_s  + \int_0^t  \int_{\cI} \sum_{i=1}^{k-1} c(s,T_i,x;y) \mu^L(ds,dy) \Bigg).
\end{align*}
By It\^o's formula for semimartingales
\begin{align}
(2') &= F(t-,T_k,x)\Bigg( \Big( \sum_{i=1}^{k-1}a(t,T_i,x)  + \frac{1}{2} \parallel \sum_{i=1}^{k-1}b(t,T_i,x) \parallel^2 \Big) dt \nonumber\\
& + \sum_{i=1}^{k-1}b(t,T_i,x) dW_t   + \int_{\er^d} \Big(e^{\la \sum_{i=1}^{k-1}b(t,T_i,x), y \ra}-1 \Big) (\mu - \nu)(dt,dy) \nonumber\\
& + \int_{\er^d} \Big(e^{\la \sum_{i=1}^{k-1}b(t,T_i,x), y \ra}-1 - \Big\la \sum_{i=1}^{k-1}b(t,T_i,x), y \Big\ra\Big)  \nu(dt,dy) \nonumber\\
&+ \int_{\cI} \Big( e^{\sum_{i=1}^{k-1}  c(t,T_i,x;y)}-1 \Big) \mu^L(dt,dy) \Bigg) \label{dyn:G}.
\end{align}
We finally incorporate the dynamics of the driving processes under the $T_k$-forward measure and obtain by
\eqref{eq:brownian-CDO} and \eqref{eq:compensator-CDO}
\begin{align*}
(2') &= F(t-,T_k,x)\Bigg( \Big( \sum_{i=1}^{k-1}a(t,T_i,x)  + \frac{1}{2} \parallel \sum_{i=1}^{k-1}b(t,T_i,x) \parallel^2  \nonumber\\
& + \Big\la \sum_{i=1}^{k-1}b(t,T_i,x),  \alpha(t,T_k) \Big\ra \\
& + \int_{\er^d} \Big(e^{\la \sum_{ i=1}^{k-1}b(t,T_i,x),  y \ra}-1  - \Big\la \sum_{ i=1}^{k-1}b(t,T_i,x),  y \Big\ra \beta(t, T_k, y)^{-1} \Big) F_t^{T_k}(dy) \\
&+ \int_{\cI} \Big( e^{\sum_{i=1}^{k-1}  c(t,T_i,x;y)}-1 \Big) F_t^{L, T_k}(dy)\Big) dt \\
& + \sum_{i=1}^{k-1}b(t,T_i,x) dW^{T_k}_t   + \int_{\er^d} \Big(e^{\sum_{i=1}^{k-1}b(t,T_i,x) y}-1 \Big) (\mu-\nu^{T_k})(dt,dy) \nonumber\\
&+ \int_{\cI} \Big( e^{\sum_{i=1}^{k-1}  c(t,T_i,x;y)}-1 \Big) (\mu^L-\nu^{L, T_k})(\dt,dy) \Bigg).
\end{align*}

It remains to calculate the covariation part $(3')$. Since $\ifL{t}{x}$ does not have a continuous martingale part, we conclude
$$
\left[G,  \ifL{\cdot}{x} \right]_{t} = \sum_{s \leq t} \Delta G(s)\Delta \ifL{s}{x}.
$$
Moreover,
\begin{align*}
\Delta \ifL{s}{x} (\omega) & = \ifL{s}{x}(\omega) - \ifL{s-}{x}(\omega)\\
\\
& = \left\{ \begin{array}{r l}
                -1 ; &  \textrm{if} \ L_{s-}(\omega) \leq x \ \textrm{and} \ L_s(\omega) > x \\
                \\
                0; & \textrm{otherwise} \, .
                \end{array}
                \right.
\end{align*}
Therefore,
\begin{eqnarray*}
\Delta \ifL{s}{x}  & = & - \indik_{\set{L_{s-} \leq x, \, L_s > x}} =   - \indik_{\set{L_{s-} \leq x, \, L_{s-} + \Delta L_s > x}}
\end{eqnarray*}
and it follows
$$
\Delta \ifL{s}{x}  = \int_{\er} z \, \mu^{\ifL{\cdot}{x}} (\set{s}, \dz) = - \int_{\cI} \ifL{s-}{x} \indik_{\set{L_{s-} + y > x}} \mu^{L} (\set{s}, \dy).
$$
In \eqref{dyn:G} we already computed the dynamics of $G$ and hence,  we deduce
\begin{eqnarray*}
(3') & = & -G(t-) \ifL{t-}{x} \int_{\cI} \Big( e^{\sum_{i=1}^{k-1}  c(t,T_i,x;y)}-1 \Big)
 \indik_{\set{L_{t-} + y > x }} \mu^L (\dt, \dy).
\end{eqnarray*}
Summing up the calculations, we obtain on $\{F(t-,T_k,x)>0\}$
\begin{align*}
\frac{dF(t,T_k,x)}{F(t-,T_k,x)} &=  \Bigg( - \lambda^{T_k}(t,x) + D(t, T_k, x)\\
&+ \int_{\cI} \Big( e^{\sum_{i=1}^{k-1}  c(t,T_i,x;y)}-1 \Big) F^{L, T_k}_t(dy) \\
&-\int_{\cI} \Big( e^{\sum_{i=1}^{k-1}  c(t,T_i,x;y)}-1 \Big)
 \indik_{\sset{L_{t-} + y > x }} F^{L, T_k}_t(dy) \Bigg) dt + d\tilde M_t,
\end{align*}
for some local martingale $\tilde M$ and with $D(t, T_k, x)$ given by \eqref{eq:D}. This concludes the proof.
\end{proof}

\begin{remark}
If the driving process $X$ does not have a Brownian part $W$, cf. \eqref{eq:driving_process}, then an inspection of the proof shows that the model is free of arbitrage if the drift condition \eqref{eq:drift_condition} holds when
the term $D(t, T_k, x)$ is replaced by
\begin{align}
\label{eq:D-pure-jump}
D(t, T_k, x ) & = \sum_{i=1}^{k-1}a(t,T_i,x)\\
& + \int_{\er^d} \Big(e^{\la \sum_{ i=1}^{k-1}b(t,T_i,x),  y \ra}-1  - \Big\la \sum_{ i=1}^{k-1}b(t,T_i,x),  y \Big\ra \beta(t, T_k, y)^{-1} \Big) F_t^{T_k}(dy) \nonumber.
\end{align}
\end{remark}

\section{Examples}\label{sec:examples}
Up to now we defined the basic ingredients for specifying models with  discrete tenor structure which are free of arbitrage.
Note that these models can be calibrated to any given initial term structure. However, for a given
family of intensities $(\lambda(t, x))_{t\geq0, x\in \cI}$ the drift has to satisfy condition \eqref{eq:drift_condition}. We shall now discuss some simple
examples which already show the high degree of flexibility. Let us repeat that this is not the case in the HJM framework developed in \citeN{FilipovicOverbeckSchmidt11} since the risky short rate in fact determines the form of the compensator of the loss process, see equation (5.1) in \citeN{FilipovicOverbeckSchmidt11}.

We start with any initial term structure, represented by a family
$H(0,T_k,x)$ for $k=0,\dots,n-1$, $x \in \cI$  and \emph{arbitrary} intensities
$(\lambda(t,x))_{t \ge 0,\ x \in \cI}$.

In the following examples we consider the case with constant term structure, see Remark \ref{rem:constantriskfreeinterest}. In this case the $T_k$-forward measures coincide and hence  $\lambda^{T_k}(t,x)=\lambda(t,x)$, $\alpha(t,T_k)=0$, $\beta(t,T_k,y)=1$, $F_t^{T_k}(dy) = F_t(dy)$ and $F_t^{L, T_k}(dy) = F_t^{L}(dy)$.
\begin{example}[Gaussian spread movements]\label{ex:Gaussian}
This example will specify a simple $d$-factor Gaussian model. We consider no jumps in
the spreads, i.e.\ $F_t(dy)=0$ and $c=0$ (no direct contagion). The volatilities $b(t,T_i,x)$
can be chosen arbitrarily, such that $(\mathbf{A2})$ is satisfied. Thereafter we proceed iteratively:
\begin{enumerate}
\item Let
$$ a(t,T_1,x) = \lambda(t,x) - \frac{1}{2}  \|b(t,T_1,x)\|^2. $$
\item For $k=2,\dots,n-1$ let
$$ a(t,T_{k},x) = \frac{1}{2}\Big( \|\sum_{i=1}^{k-1} b(t,T_i,x)\|^2 - \|\sum_{i=1}^{k} b(t,T_i,x)\|^2\Big). $$
\end{enumerate}
Clearly, this model is free of arbitrage and can be calibrated to any given initial term structure.
Note that the drift of the
$H$ with closest maturity compensates the intensity $\lambda(t,x)$.
\end{example}

\begin{example}[L\'evy driven spread movements without Gaussian component]\label{ex:Levy}
We assume pure-jump spread movements such that \eqref{eq:D-pure-jump} holds. With $c=0$, we proceed analogously to
the Gaussian example and start with arbitrary $F_t(dy)$ and $b(t,T_i,x)$  such that $(\mathbf{A1})$ and $(\mathbf{A2})$ are satisfied.
\begin{enumerate}
\item Define
$$ a(t,T_1,x) = \lambda(t,x) -
 \int_{\er^d} \Big(e^{\la b(t,T_1,x),  y \ra}-1  - \la b(t,T_1,x),  y \ra \Big) F_t(dy). $$
\item For $k=2,\dots,n-1$ define
\begin{align*} a(t,T_{k},x) =
 \int_{\er^d} \Big(e^{\la \sum_{ i=1}^{k-1}b(t,T_i,x),  y \ra}-1  - \Big\la \sum_{ i=1}^{k-1}b(t,T_i,x),  y \Big\ra \Big) F_t(dy) \\
- \int_{\er^d} \Big(e^{\la \sum_{ i=1}^{k}b(t,T_i,x),  y \ra}-1  - \Big\la \sum_{ i=1}^{k}b(t,T_i,x),  y \Big\ra \Big) F_t(dy).
\end{align*}
\end{enumerate}
\end{example}

\begin{example}[Contagion] Next, we incorporate a direct contagion, i.e.\ $c$ does not vanish.
We continue with the L\'evy setting of Example \ref{ex:Levy}.
Contagion can be specified via the function $c$: if the loss process has a jump of size $y$ at $t$, then
$$
H(t,T_k,x) = H(t-,T_k,x) e^{c(t,T_k,x;y)}
$$
since $X$ and $L$ do not jump simultaneously. We can specify an arbitrage-free model with the following steps.
\begin{enumerate}
\item Let
\begin{align*}
a(t,T_1,x) &= \lambda(t,x) -
 \int_{\er^d} \Big(e^{\la b(t,T_1,x),  y \ra}-1  - \la b(t,T_1,x),  y  \ra \Big) F_t(dy) \\
 &-  \int_{\cI} \Big( e^{c(t,T_1,x;y)} -1 \Big)\ind{L_{t-}+y \le x} F_t^L(dy).
\end{align*}
\item For $k=2,\dots,n-1$ let
\begin{align*} a(t,T_{k},x) &=
 \int_{\er^d} \Big(e^{\la \sum_{ i=1}^{k-1}b(t,T_i,x),  y \ra}-1  - \Big\la \sum_{ i=1}^{k-1}b(t,T_i,x),  y  \Big\ra \Big) F_t(dy) \\
  &+  \int_{\cI} \Big( e^{\sum_{i=1}^{k-1}c(t,T_1,x;y)} -1 \Big)\ind{L_{t-}+y \le x} F_t^L(dy) \\
&- \int_{\er^d} \Big(e^{\la \sum_{ i=1}^{k}b(t,T_i,x),  y \ra}-1  - \Big\la \sum_{ i=1}^{k}b(t,T_i,x),  y \Big\ra \Big) F_t(dy)
 \\
  &-  \int_{\cI} \Big( e^{\sum_{i=1}^k c(t,T_i,x;y)} -1 \Big)\ind{L_{t-}+y \le x} F_t^L(dy).
\end{align*}
\end{enumerate}
\end{example}

For some applications it may be interesting to simplify this setting further.
As examples we discuss \emph{additive} and \emph{multiplicative} jumps in $H$.

\begin{enumerate}
\item \emph{Additive jumps.}
We choose (deterministic) functions $C(t,x)$ and let
$$
 e^{c(t,T_k,x;y)}  := H(t-, T_k, x)^{-1} y \, C(T_k-t,x) + 1.
$$
This yields a jump of size $\Delta L_t C(T_k-t,x)$ of $H$ at time $t$, i.e.
$$ H(t,T_k,x) = H(t-,T_k,x) + \Delta L_t C(T_k-t,x),   $$
while the specification
$$
e^{c(t,T_k,x;y)}:=\big( 1 + H(t-, T_k, x) y \, C(T_k-t,x) \big)^{-1}
$$
yields a jump of size $\delta_k^{-1} \Delta L_t C(T_k-t,x)$ in the credit spread as defined in formula \eqref{eq:creditspread}:
$$
cs(t,T_k,x) = cs(t-,T_k,x) + \delta_k^{-1} \Delta L_t C(T_k-t,x).
$$
\item \emph{Multiplicative jumps.}
Again we choose (deterministic) functions $C(t,x)$ and let
$$
e^{c(t,T_k,x;y)}:= y \, C(T_k-t,x).
$$
In this case,
$$
H(t,T_k,x) = H(t-, T_k, x) \Delta L_t \, C(T_k-t,x)
$$
and in the drift condition we have the following simplification
\begin{align}\label{ex:levycontation1}
\lefteqn{\int_{\cI} \Big(e^{\sum_{i=1}^{k-1} c(t,T_i,x;y)} -1 \Big)\ind{L_{t-}+y \le x} F_t^L(dy) } \hspace{2cm} \nonumber\\
&= \int_{\cI} \Big( y^{k-1}  \prod_{i=1}^{k-1} C(T_i-t,x) -1 \Big)\ind{L_{t-}+y \le x} F_t^L(dy).
\end{align}
This expression depends on the distribution of the losses via $F_t^L$.
For various approaches concerning the dependence on the loss process see  \citeN{ContDeguestKan10}.
\end{enumerate}

\begin{example}[Relation to a bottom-up model]
Continuing Remark \ref{rem:bottom-up} we consider
 a bottom-up model with $m$ entities and associated default times $\tau_1,\dots,\tau_m$. The loss process is
$$ L_t = \sum_{i=1}^m \ind{\tau_i \le t} q_i, $$
where $q_i$ is the loss given default of entity $i$. { Assume that $q_i$  are constant and
$\tau_i$ has default intensity $\lambda_i$, that is
$$ \ind{\tau_i \le t} - \int_0^{t}\ind{\tau_i > s} \lambda_i(s) ds $$
is a martingale for $i=1,\dots,m$. Then
the compensator of $L$ is
$$ \nu^L(dt,dx) = F_t^L(dx)dt = \sum_{i=1}^m \lambda_i(t) \ind{\tau_i > t} \delta_{\{q_i\}}(dx)dt. $$
For intuition consider i.i.d.\ exponentially distributed $\tau_i$  where the intensity parameter is $\lambda$
and $q_i=q.$ Then
$$ F_t^L(dx) = \lambda \sum_{i=1}^m \ind{\tau_i > t} \delta_{\{q\}} (dx) = \lambda (m- q^{-1} L_{t}) \delta_{\{q\}} (dx). $$}
Note that the compensator naturally depends on the number of defaults that have occurred already: as less and less entities
remain in the pool, the intensity for a further loss decreases.
\end{example}

%%%%%%%%%%%%%%%%%%%%%%%%%%%%%%%%%%%%%%%%%%%%%%%%%%%%%%%%%%%%%%%%%%%%%%%%%%%%%%%%%%%%%%
%
% An affine specification
%
%%%%%%%%%%%%%%%%%%%%%%%%%%%%%%%%%%%%%%%%%%%%%%%%%%%%%%%%%%%%%%%%%%%%%%%%%%%%%%%%%%%%%%

\section{An affine specification}
\label{sec:affine}
Affine processes are a powerful tool for yield curve modeling because they represent a rich class of
processes, allowing for jumps and stochastic volatility, while still retaining a high
degree of tractability. For examples see \citeN{CuchieroFilipovicTeichmann09},
and \citeN{ErraisGieseckeGoldberg10} for self-exciting affine processes. \citeN{DuffieGarleanu01} is to our knowledge the first paper using affine jump-diffusions for modeling of stochastic intensities of single obligors in a dynamic bottom-up credit portfolio model. This section will illustrate how these processes can be used in our setup.
Note that this is very different from the setting in \citeN{FilipovicOverbeckSchmidt11};
already Example \ref{ex:Gaussian} illustrates that Gaussian behavior of the spreads in a  model with discrete tenor structure is possible,
while in their setting this would generate arbitrage possibilities, see also Remark \ref{rem:affineFOS}.
Moreover, in our approach we are able to find an affine specification which includes contagion
as we will show in the following.

For simplicity we discuss only the case of affine processes which are driven by a diffu\-sion and a
constant term structure as in Remark \ref{rem:constantriskfreeinterest}.
Denote by $\cT:=\{T_0,\dots,T_n\}$ the tenor structure and
let $\cZ\subset \R^d$ be some closed state space with nonempty interior.
Consider a $d$-dimensional Brownian motion $W$ and let $\mu$ be defined on $\cZ$ by
  $$
   \mu(z) = \mu_0 + \sum_{i=1}^d\mu_i\, z_i,
  $$
  for some vectors $\mu_i \in \R^d$, $i=0,\dots,d$. Furthermore, we assume that $\sigma$ is defined on $\cZ$ with values in $\R^{d\times d}$ such that
  \begin{align}\label{eq:sigma}
  \frac{1}{2} \sigma(z)^{\top}\sigma(z) = \nu_0 + \sum_{i=1}^d \nu_i \, z_i,
  \end{align}
  for some matrices $\nu_i \in \R^{d\times d}$, $i=0,\dots,d$.
For any $z \in \cZ$ we denote by $Z=Z^z$ the continuous, unique strong solution of
$$ dZ_t = \mu(Z_t) dt + \sigma(Z_t) dW_t, \quad Z_0=z. $$
The class of models we consider are of the form
\begin{align} \label{eq:Haffine}
H(t,T_k,x) = \exp\bigg( & A(t,T_k,x) + B(t,T_k,x)^\top Z_t  \\
 &+ \int_0^t\int_I c(s,T_k,x,L_{s-};y) \mu^L(ds,dy) + \int_0^t d(s, T_k, x,L_{s-},Z_s)ds\bigg). \nonumber
\end{align}
The first line is the part which is affine while the second part considers a contagion term
which can have arbitrary dependence on $L$, but no dependence on $Z$.
The term $d$ defines a drift which will compensate default and contagion risk.
The assumptions on the functions $A$, $B$, $c$, and $d$ are as follows:
\begin{itemize}
\item[$(\mathbf{B1})$] $A$ and $B$ satisfy the following system of Riccati equations:
\begin{align}
-\partial_t A(t,T_k,x) &= B(t,T_k,x)^\top \mu_0 + B(t,T_k,x)^\top 2 \nu_0 \sum_{i=1}^k B(t,T_i,x)
\label{eq:affineA}\\
&-B(t,T_k,x)^\top \nu_0 B(t,T_k,x), \nonumber\\
-\partial_t B(t,T_k,x)_j &= B(t,T_k,x)^\top \mu_j + B(t,T_k,x)^\top 2\nu_j \sum_{i=1}^k B(t,T_i,x)
\label{eq:affineB}\\
&-B(t,T_k,x)^\top \nu_jB(t,T_k,x), \nonumber
\end{align}
for $0 \le t \le T_k$.
\item[$(\mathbf{B2})$] The function $c:\R^+\times\cT\times\cI\times\cI\times\cI$ satisfies
$$\sup_{t \le T_k,x,l,y\in\cI} |c(t,T_k,x,l;y)| < \infty$$
\item[$(\mathbf{B3})$] The compensator of the loss process satisfies $F_t^L(A) = m(t,L_{t-},Z_t,A)$
for all $A\in\cB(\cI)$ where $m(t,l,z,\cdot)$ is a $\sigma$-finite Borel measure for each
$(t,l,z)\in\R^+\times\cI\times\cZ$. Moreover $m$ is affine, i.e.
$$ m(t,l,z,\cdot) = m_0(t,l,\cdot) + \sum_{i=1}^d m_i(t,l,\cdot)z_i $$
for some $m_i:\R^+\times\cI\times\cB(\cI)\to\R^+$, $i=0,\dots,d$.
\item[$(\mathbf{B4})$] The additional drift is affine, i.e.\
$$ d(t,T_k,x,l,z) = d_0(t,T_k,x,l) + \sum_{i=1}^d d_i(t,T_k,x,l) z_i, \quad k=1,\dots,n  $$
and
\begin{align*}
d_i(t,T_1,x,l) &= \int_{\cI}\Big(1-e^{c(t,T_1,x,l;y)} \ind{y \le x-l}\Big) m_i(t,l,dy) \\
d_i(t,T_k,x,l) &= \int_{\cI}\Big(e^{\sum_{j=1}^{k-1}c(t,T_j,x,l;y)} - e^{\sum_{j=1}^{k}c(t,T_j,x,l;y)} \Big)
\ind{y \le x-l} m_i(t,l,dy)
\end{align*}
for $i=0,\dots,d$ and $k=2,\dots,n$.
\end{itemize}
\begin{remark}
Note that in ({\bf B3}) we require  $m(t,l,z,\cdot)$ not to be a signed measure. This implies
restrictions on $m_i$ depending on the state space: if $\cZ=\R^{d_1}\times(\R^+)^{d_2}$, with $d_1>0$ and $d=d_1+d_2$,
then $m_i(t,l,\cdot)=0$ for $i=1,\dots,d_1$ as otherwise there exist $z\in\cZ$ such that
$$ m_0(t,l,A) + \sum_{i=1}^{d} m_i(t,l,A)z_i <0 $$
for some $l$ and $A$. This contradicts $F_t^L(A) = m(t,L_{t-},Z_t,A)\ge 0$.
\end{remark}

We assume that all functions which appear here are c\`adl\`ag  in each variable.

The input parameters for the model are the coefficients $\mu_i$, $\nu_i$, as well as the
contagion function $c$ and the Borel-measures $m_i$, $i=0,\dots,d$.
Note that we do not need to specify boundary conditions on the Riccati equations.
They can be used to improve the fit on the initial term structure.
The following proposition
shows that the above conditions lead indeed to an arbitrage-free model.

\begin{proposition} \label{prop:affine}
Assume  $(\mathbf{B1})$-$(\mathbf{B4})$. Then $(F(t,T_k,x))_{0 \le t \le T_{k-1}}$ given by \eqref{def:F1}
with $H$ as in  \eqref{eq:Haffine} are $\Q^*$-local martingales.
\end{proposition}

We start with a small lemma which is proved directly by applying It\^o's formula.
\begin{lemma}
Consider $H$ as in \eqref{eq:Haffine} and assume that $A$ and $B$ are differentiable in $t$
with c\`adl\`ag  derivatives. Then $H$ can be represented as in \eqref{def:H} with
\begin{align*}
a(t,T_k,x) &= \partial_t A(t,T_k,x)+ \partial_t B(t,T_k,x)^\top Z_t + B(t,T_k,x)^\top\mu(Z_t) \\
&+d(t,T_k,x,L_{t-},Z_t) \\
b(t,T_k,x) &= B(t,T_k,x)^\top \sigma(Z_t)\\
c(t,T_k,x;y) &= c(t,T_k,x,L_{t-};y).
\end{align*}
\end{lemma}

\begin{proofprop}{\ref{prop:affine}}
Note that all assumptions of Theorem \ref{th:drift-cond} are satisfied. In particular $(\mathbf{A1})$
is trivially true since $F_t$ is $0$ as a consequence of the continuity of $(Z_t)$. At the same time
this allows to choose $\tilde C$ in $(\mathbf{A1})$ equal to infinity and $(\mathbf{A2})$ follows.

Our aim is to show that the drift condition \eqref{eq:drift_condition} is satisfied. In this regard, consider the case where $X$ is the $d$-dimensional Brownian motion $W$. We compute
\begin{align}
&\sum_{i=1}^{k-1}a(t,T_i,x)
+ \frac{1}{2} \parallel \sum_{i=1}^{k-1}b(t,T_i,x) \parallel^2 \nonumber \\
&+ \int_{\cI} \Big( e^{\sum_{i=1}^{k-1}  c(t,T_i,x;y)}-1 \Big) \indik_{\sset{L_{t-} + y \le x }} F^{L}_t(dy)
-\lambda(t,x) \nonumber\\
&\qquad= \sum_{i=1}^{k-1} \Big(\partial_t A(t,T_k,x)+ \partial_t B(t,T_k,x)^\top Z_t + B(t,T_k,x)^\top\mu(Z_t) \Big) \nonumber \\
 &\qquad+ \frac{1}{2} \parallel \sum_{i=1}^{k-1} B(t,T_i,x)^\top \sigma(Z_t) \parallel^2 \nonumber\\
 &\qquad+ \sum_{i=1}^{k-1} d(t,T_i,x,L_{t-},Z_t) \label{ea:2}\\
& \qquad+ \int_{\cI} \Big( e^{\sum_{i=1}^{k-1}  c(t,T_i,x,L_{t-};y)}-1 \Big) \indik_{\sset{L_{t-} + y \le x }}
 m(t,L_{t-},Z_t,dy)-\lambda(t,x) .\label{ea:3}
 \end{align}
Note that according to \eqref{eqn:lambdatx} $\lambda(t,x)=m(t,L_{t-},Z_t,(x-L_{t},1]\cap \cI)$.
Now we consider the equation above for all possible values $l\in\cI$ of $L_t$ and $z\in\cZ$ of $Z_t$.
We have that $m(t,l,z,[0,x-l]\cap\cI)+m(t,l,z,(x-l,1]\cap\cI) = m(t,l,z,\cI)$ and we obtain
\begin{align*}
\eqref{ea:3} &=
 \int_{\cI} e^{\sum_{i=1}^{k-1}  c(t,T_i,x,l;y)}  \indik_{\sset{l + y \le x }}  m(t,l,z,dy) - m(t,l,z,\cI).
\end{align*}
We set $z_0\equiv1$ to simplify the notation. By $(\mathbf{B4})$, we obtain
\begin{align*}
\eqref{ea:2} &= d(t,T_1,x,l,z)+\sum_{i=2}^{k-1}d(t,T_i,x,l,z) \\
&= \sum_{j=0}^d z_j \bigg(\int_{\cI}\Big(1-e^{c(t,T_1,x,l;y)} \ind{y \le x-l}\Big) m_j(t,l,dy)     \\
&\qquad + \sum_{i=2}^{k-1}\int_{\cI}\Big(e^{\sum_{j'=1}^{i-1}c(t,T_{j'},x,l;y)} -
e^{\sum_{j'=1}^{i}c(t,T_{j'},x,l;y)} \Big)
\ind{y \le x-l} m_j(t,l,dy) \bigg) \\
&= \sum_{j=0}^d z_j \bigg(\int_{\cI}\Big(1 - e^{\sum_{j'=1}^{k-1}c(t,T_{j'},x,l;y)}
\ind{y \le x-l}  \Big)m_j(t,l,dy) \bigg) \\
&= \int_{\cI}\Big(1 - e^{\sum_{j'=1}^{k-1}c(t,T_{j'},x,l;y)}
\ind{y \le x-l}\Big) m(t,l,z,dy).
\end{align*}
Hence, $\eqref{ea:2}+\eqref{ea:3}=0$. Our final step consists in proving that
\begin{align}\label{temp1061}
 0& = \sum_{i=1}^{k-1}\Big(\partial_t A(t,T_i,x)+ \partial_t B(t,T_i,x)^\top z +
B(t,T_i,x)^\top\mu(z) \Big) \nonumber \\
&+ \frac{1}{2} \parallel \sum_{i=1}^{k-1} B(t,T_i,x)^\top \sigma(z) \parallel^2.
\end{align}
As this equation is affine in $z$, i.e.\ of the form $\sum_{i=0}^d \alpha_i z_i$, it
is sufficient to show that $\alpha_i=0$ for $i=0,\dots,d$. First, we consider $\alpha_0$ and
show that
\begin{align}\label{temp1068}
0 = \sum_{i=1}^{k-1}\Big(\partial_t A(t,T_i,x) + B(t,T_i,x)^\top \mu_0\Big) +
\sum_{i,j=1}^{k-1}B(t,T_i,x)^\top \nu_0 B(t,T_j,x).
\end{align}
Note that \eqref{eq:sigma} implies that   $\nu_j$ is symmetric for any $j=1,\dots,d$.
Hence, by $(\mathbf{B1})$,
\begin{align*}
0&=\sum_{i=1}^{k-1} \Big(\partial_t A(t,T_i,x) + B(t,T_i,x)^\top \mu_0\Big) \\
&+\sum_{i=1}^{k-1}B(t,T_i,x)^\top \nu_0 \sum_{j=1}^{i} B(t,T_j,x) \\
&+\sum_{i=1}^{k-1}B(t,T_i,x)^\top \nu_0 \sum_{j=1}^{i} B(t,T_j,x) \\
&- \sum_{i=1}^{k-1}B(t,T_i,x)^\top \nu_0 B(t,T_i,x)
\end{align*}
an this is exactly  \eqref{temp1068}. In a similar way, $(\mathbf{B1})$ yields
\begin{align*}
0=\sum_{i=1}^{k-1}\Big(\partial_t B(t,T_i,x)_j + B(t,T_i,x)^\top \mu_j\Big) +
\sum_{i,l=1}^{k-1}B(t,T_i,x)^\top \nu_j B(t,T_l,x)
\end{align*}
for $j=1,\dots,d$ such that \eqref{temp1061} is proven.
Summarizing, we obtain that the drift condition \eqref{eq:drift_condition} holds
and we conclude by Theorem \ref{th:drift-cond}.
\end{proofprop}

\begin{remark}
The previous proof shows that the coupled Riccati equations for $A$ and $B$ may be simplified by considering
$$  A^k(t,x) := \sum_{i=1}^k A(t,T_i,x), \quad B^k(t,x) := \sum_{i=1}^k B(t,T_i,x). $$
Then \eqref{eq:affineA} and \eqref{eq:affineB} are equivalent to
\begin{align}
-\partial_t  A^k(t,x) &=  B^k(t,x)\mu_0 +  B^k(t,x)^\top \nu_0  B^k(t,x)
\label{eq:affineA2}\\[1ex]
-\partial_t B^k(t,x)_j &= B^k(t,x)\mu_j +  B^k(t,x)^\top \nu_j  B^k(t,x)
\label{eq:affineB2}
\end{align}
for $k=1,\dots,n$ and $j=1,\dots,d$.
Equations \eqref{eq:affineA2} and \eqref{eq:affineB2} are the classical Riccati equations for multivariate affine processes.
In dimension $d=1$ the solutions are well-known, while in the general case
efficient numerical schemes are available to compute $A^k$ and $B^k$.
\end{remark}

Up to now the modeling was quite general. In the following example
we give a concrete one-dimensional affine specification which is much simpler.
We will use a two-dimensional extension later on in the section on calibration.

\begin{example} We choose a Feller square-root process as a driver: consider $d=1$ and $\mu_0 \ge 0,$ $\mu_1 \in \R$ as well as $\nu_1 = \sigma^2/2$. Then
\begin{align*}
dZ_t&= (\mu_0 + \mu_1  Z_t )dt + \sigma\sqrt{Z_t} dW_t,
\end{align*}
with $Z_0=z>0$. The Feller condition $2 \mu_1 > \sigma^2$ ensures positivity of $Z$.
In this case the Riccati equations \eqref{eq:affineA2} and \eqref{eq:affineB2} have
explicit solutions, see for example \citeN{CuchieroFilipovicTeichmann09}.
The compensator of the loss process is specified via
$$ m(t,l, z, dy) = m_0 + m_1 p_{\alpha,\beta}(dy) z, $$
where $p_{\alpha,\beta}$ is a Beta$(\alpha,\beta)$-distribution. Finally,
the contagion parameter is assumed to be a function of the loss process, i.e.
$$ c(t,T_k,x,l;y) = c(T_k-t,y). $$
Choosing $c$ decreasing in $y$ guarantees that upward jumps in the loss process
lead to downward jumps in the price process, and hence
to upward jumps in the credit spreads. Computing the terms $d_1,\dots,d_k$
by a simple numerical integration is the last step for specifying an arbitrage-free model.
\end{example}

%%%%%%%%%%%%%%%%%%%%%%%%%%%%%%%%%%%%%%%%%%%%%%%%%%%%%%%%%%%%%%%%%%%%%%%%%%%%%%%%%%%%%%
%
% Pricing of STCDOS
%
%%%%%%%%%%%%%%%%%%%%%%%%%%%%%%%%%%%%%%%%%%%%%%%%%%%%%%%%%%%%%%%%%%%%%%%%%%%%%%%%%%%%%%

\section{Pricing of portfolio credit derivatives}
\label{sec:pricing}
In this section we study the valuation of portfolio credit derivatives. In particular, we focus our attention on single tranche CDOs (STCDOs) and call options on STCDOs.
\subsection{Single tranche CDO} The  valuation of derivatives can often be facilitated by using appropriate defaultable forward measures. We illustrate this by considering a standard instrument for investment in a  credit pool, a so-called single tranche CDO.
A \emph{single tranche CDO} (STCDO)  is specified by:
\begin{itemize}
\item[-] a collection of future dates (tenor dates) $T_{1} < T_{2} < \cdots < T_{m}$,
\item[-]  \emph{lower} and \emph{upper detachment points}
$x_{1}<x_{2}$ in $[0,1]$
\item[-] {a fixed spread} $S$.
\end{itemize}
The STCDO offers premium in exchanges for payments at defaults: the \emph{premium leg} (received by the investor) consists of a series of payments equal to
\begin{align}\label{eq:S-STCDO}
S [(x_{2}-L_{T_{k}})^{+} - (x_{1}-L_{T_{k}})^{+}],
\end{align}
received at $T_{k}$, $k=1, \ldots, m-1$.
Letting
\begin{align}
\label{eq:f}
f(x)& :=(x_{2}-x)^{+} - (x_{1}-x)^{+} = \int_{x_{1}}^{x_{2}} \indik_{\set{x \leq y}} \dy,
\end{align}
we have that $\eqref{eq:S-STCDO}=Sf(L_{T_k}). $

The \emph{default leg}  (paid by the investor) consists of a series of payments at times $T_{k+1}$,  $k=1, \ldots, m-1$, given by
\begin{equation}
\label{eq:default_leg_payment}
f(L_{T_{k}}) - f(L_{T_{k+1}}).
\end{equation}
This payment is nonzero only if $\Delta L_{t} \neq 0$ for some $t \in (T_{k}, T_{k+1}]$. In the literature alternative payment schemes can be found as well (see \citeN{FilipovicOverbeckSchmidt11}, for example).
We have
\begin{eqnarray*}
(\ref{eq:default_leg_payment})  & = & \int_{x_{1}}^{x_{2}} \Big[\ifL{T_k}{y} - \ifL{T_{k+1}}{y} \Big]\dy
= \int_{x_{1}}^{x_{2}} \indik_{\sset{L_{T_{k}} \leq y, L_{T_{k+1}} > y}} \dy.
\end{eqnarray*}
Let us denote  by $e(t, T_{k+1}, x)$ the value at time $t$ of a payment given by $\indik_{\{L_{T_{k}} \leq x, L_{T_{k+1}} > x\}}$ at the tenor date $T_{k+1}$.
To calculate $e(t, T_{k+1}, x)$, it is convenient to replace the measure $\Q_{T_{k+1}}$ by a new one.
As already discussed, the market trades only financial instruments whose first tenor date is at least a full
tenor period away. In this regard we introduce a time horizon $\delta<T_1$ and consider
the forward prices on $[0,\delta]$. Applying Theorem \ref{th:drift-cond}
with respect to the  tenor structure $\{\delta,T_1,\dots,T_m\}$
yields an arbitrage-free construction of forward prices.
Assume
\begin{itemize}
\item[$(\mathbf{A5})$]
The processes $(F(t,T_k,x))_{0 \leq t \leq T_{k-1}}$, are true $\Q_{T_k}$-martingales
for every $k=2,\dots,n$ and $x \in \cI$. Moreover, $(F(t,T_1,x))_{0 \leq t \le \delta}$
is a true $\Q_{T_1}$-martingale.
\end{itemize}

\noindent Assumption $(\mathbf{A5})$ allows us to switch  to a measure under which the numeraire is given by the $(T_k,x)$-forward price. This is not
an equivalent measure change, but it still yields a measure which is absolutely continuous with respect to the initial one.
 Similar measure changes have been introduced in \citeN{Schoenbucher00b} and have been successfully applied to the pricing of credit risky securities, cf. \citeN{EberleinKlugeSchoenbucher06}.
Let $x \in [0, 1]$ and $k \in \set{1, \ldots, m-1}$. We define the  $(T_{k+1},x)$-\emph{forward measure} $\Q_{T_{k+1},x}$ on $(\Omega, \cG_{T_{k+1}})$ by its Radon--Nikodym derivative
\begin{eqnarray*}
\frac{\ud \Q_{T_{k+1},x}}{ \ud \Q_{T_{k+1}}} & := & \frac{F(T_k,T_{k+1},x)}{\E_{\Q_{T_{k+1}}}\left[ F(T_k,T_{k+1},x)\right]}
=  \frac{F(T_k,T_{k+1},x)}{F(0,T_{k+1},x)},
\end{eqnarray*}
where the last equality follows under $(\mathbf{A5})$.
The corresponding density process is
$$
\frac{\ud \Q_{T_{k+1},x}}{ \ud \Q_{T_{k+1}}}\Bigg|_{\cG_{t}} = \frac{F(t,T_{k+1},x)}{F(0,T_{k+1},x)}.
$$
As already mentioned,  $\Q_{T_{k+1},x}$ is not equivalent to $\Q_{T_{k+1}}$ if $\Q_{T_{k+1}}(L_{T_{k}}> x) >0$.

\begin{lemma}
\label{le:value_of_crossing_x}
Assume $(\mathbf{A5})$. Let $x \in \cI$ and $k \in \set{1, \ldots, m-1}$. Then,  for every $t \leq T_{k}$,
\begin{equation*}
e(t, T_{k+1}, x) = P(t, T_{k+1}, x) \E_{\Q_{T_{k+1},x}} \bigg(\prod_{i=0}^{k}H(T_i, T_i, x)^{-1} - 1| \cG_{t}\bigg).
\end{equation*}
\end{lemma}

\begin{proof}
The  price at time $t$ of a contingent claim with payoff
$$
e(T_{k+1}, T_{k+1}, x) = \ifL{T_k}{x} - \ifL{T_{k+1}}{x}
$$
at $T_{k+1}$ equals
\begin{equation}
\label{eq:expression-for-e}
e(t, T_{k+1}, x) = P(t, T_{k+1}) \E_{\Q_{T_{k+1}}}\Big(\ifL{T_k}{x}-\ifL{T_{k+1}}{x} | \cG_{t}\Big).
\end{equation}
Regarding the second term, observe that
\begin{equation}
\label{eq:time-t-bond-price}
P(t, T_{k+1}) \E_{\Q_{T_{k+1}}}\big(\ifL{T_{k+1}}{x} | \cG_{t}\big) = P(t, T_{k+1}, x)
\end{equation}
by $(\mathbf{A5})$.
For the first term we have
\begin{eqnarray*}
\lefteqn{\E_{\Q_{T_{k+1}}}\big(\ifL{T_k}{x} | \cG_{t}\big)} \\
&& \quad \quad  = \E_{\Q_{T_{k+1}}}\bigg(\ifL{T_k}{x} \left( \prod_{i=0}^{k}H(T_k, T_i, x)\right) \left(\prod_{i=0}^{k}H(T_k, T_i, x)\right)^{-1} \Big| \cG_{t}\bigg)\\
&&  \quad \quad = \E_{\Q_{T_{k+1}}}\Big(F(T_k,T_{k+1},x) \prod_{i=0}^{k}H(T_i, T_i, x)^{-1} \Big| \cG_{t}\Big),
\end{eqnarray*}
which follows from
\eqref{def:F1} and $H(t, T_i, x) = H(T_i, T_i, x)$, for $t \geq T_i$.
Changing to the measure $\Q_{T_{k+1},x}$ yields$$
\E_{\Q_{T_{k+1}}}(\ifL{T_k}{x} | \cG_{t}) = F(t, T_{k+1},x) \E_{\Q_{T_{k+1},x}}\bigg(  \prod_{i=0}^{k}H(T_i, T_i, x)^{-1} \Big| \cG_{t}\bigg).
$$
Therefore,
\begin{align*}
e(t, T_{k+1}, x) & =  P(t, T_{k+1})  \frac{P(t, T_{k+1},x)}{P(t, T_{k+1})} \E_{\Q_{T_{k+1},x}}\bigg( \prod_{i=0}^{k}H(T_i, T_i, x)^{-1} \Big| \cG_{t} \bigg) \\
\\
&  - P(t, T_{k+1}, x) \\
\\
&  =  P(t, T_{k+1}, x) \E_{\Q_{T_{k+1},x}} \bigg(\prod_{i=0}^{k}H(T_i, T_i, x)^{-1} -1 | \cG_{t}\bigg)
\end{align*}
and the lemma is proved.
\end{proof}

\begin{proposition}
\label{prop:stcdo-price}
Assume $(\mathbf{A5})$. Then
the value of the STCDO at any time $t\in[0,\delta]$ is
\begin{equation}
\label{eq:STCDO}
\pi^{STCDO}(t, S)= \int_{x_{1}}^{x_{2}} \Big( S \sum_{k=1}^{m-1} P(t, T_{k},y) - \sum_{k=1}^{m-1} e(t, T_{k+1}, y)\Big) \dy.
\end{equation}
\end{proposition}
\noindent Recall that the premium $S f(L_{T_{k}})$ is paid at times $T_{1}, \ldots, T_{m-1}$, whereas the default payments are due at time points $T_{2}, \ldots, T_{m}$. \\[-1mm]

\begin{proof}
The value of the premium leg at time $t$ equals
\begin{eqnarray*}
\sum_{k=1}^{m-1} P(t, T_{k}) \E_{\Q_{T_{k}}} (S f(L_{T_{k}}) |\cG_t ) & = & \sum_{k=1}^{m-1} S P(t, T_{k}) \int_{x_{1}}^{x_{2}} \E_{\Q_{T_{k}}} (\ifL{T_k}{y} |\cG_t ) \dy \\
& = & S \sum_{k=1}^{m-1}  \int_{x_{1}}^{x_{2}} P(t, T_{k}, y) \dy,
\end{eqnarray*}
where we have used \eqref{eq:time-t-bond-price}.
On the other side, the default payment at time $T_{k+1}$ is given by $f(L_{T_k})-f(L_{T_{k+1}})$. Its value at time $t$ is equal to
\begin{eqnarray}\label{eq:e}
\lefteqn{ P(t, T_{k+1}) \E_{\Q_{T_{k+1}}} (f(L_{T_k}) - f(L_{T_{k+1}})\,|\,\cG_t)} \\
&& \qquad = P(t, T_{k+1}) \E_{\Q_{T_{k+1}}} \bigg( \ \int_{x_1}^{x_2} \indik_{\sset{L_{T_{k}} \leq y, L_{T_{k+1}} > y}} \dy  \,\big|\,\cG_t\bigg) \nonumber\\
&& \qquad =  \int_{x_1}^{x_2} P(t, T_{k+1}) \E_{\Q_{T_{k+1}}} \left(\indik_{\sset{L_{T_{k}} \leq y, L_{T_{k+1}} > y}}
\,\big|\,\cG_t \right) \dy \nonumber\\
&&  \qquad = \int_{x_{1}}^{x_{2}} e(t, T_{k+1}, y) \dy. \nonumber
\end{eqnarray}
Hence, the value of the default leg at time $t$ is given by
$$
\sum_{k=1}^{m-1} \ \int_{x_{1}}^{x_{2}}e(t, T_{k+1}, y) \dy.
$$
Finally, the value of the STCDO is the difference of these two values and thus we obtain \eqref{eq:STCDO}.
\end{proof}

The STCDO spread $S^{*}_{t}$ at time $t$ is the spread which makes the value of the STCDO equal to zero, i.e. one has to solve $\pi^{STCDO}(t, S) = 0$. The previous proposition yields
\begin{align}\label{eq:spread}
S^{*}_{t} = \frac{\int_{x_{1}}^{x_{2}} \sum_{k=1}^{m-1} e(t, T_{k+1}, y) \dy}{ \int_{x_{1}}^{x_{2}} \sum_{k=1}^{m-1} P(t, T_{k},y) \dy}.
\end{align}

\begin{corollary}
\label{cor:cond-indep}
Assume $(\mathbf{A5})$ and assume that the default-free bond prices $P(\cdot, T_k)$ and the loss process $L$ are conditionally independent given $\cG_t$, for all $k \in \{1, \ldots, n\}$ and $t \in [0, \delta]$. Then
\begin{equation}
\label{eq:STCDO-cond-indep}
e(t, T_{k+1}, x) = P(t, T_{k+1}) F(t, T_k, x) - P(t, T_{k+1}, x).
\end{equation}
\end{corollary}

\begin{proof}
Conditional independence of $P(\cdot, T_k)$ and $L$ implies
$$
\E_{\Q_{T_{k+1}}} ( \indik_{\{L_{T_k} \le x\}}| \cG_t) = \E_{\Q_{T_{k}}} ( \indik_{\{L_{T_k} \le x\}}| \cG_t),
$$
since $\frac{\ud \Q_{T_{k}}}{\ud \Q_{T_{k+1}}} |_{\cG_t} = \frac{P(0, T_{k+1})}{P(0, T_k)} \frac{P(t, T_k)}{P(t, T_{k+1})}$ is the density process for this change of measure (cf. equation \eqref{eq:measure-change-default-free-CDO}). Then
$$
\E_{\Q_{T_{k+1}}} ( \indik_{\{L_{T_k} \le x\}}| \cG_t) = \frac{P(t, T_k, x)}{P(t, T_k)} = F(t, T_k, x)
$$
and we obtain from \eqref{eq:expression-for-e}
$$
e(t, T_{k+1}, x) = P(t, T_{k+1}) F(t, T_k, x) - P(t, T_{k+1}, x).
$$
\end{proof}

\begin{corollary}
\label{cor:STCDO-cond-indep}
Under the assumptions of Corollary \ref{cor:cond-indep}, the price at time $t\in[0,\delta]$ of the STCDO is given by
\begin{equation}
\label{eq:STCDO-price-cond-indep}
\pi^{STCDO}(t, S) = \int_{x_1}^{x_2} \left( \sum_{k=1}^{m} c_k P(t, T_k) F(t, T_k, y) - \sum_{k=1}^{m-1} P(t, T_{k+1}) F(t, T_k, y) \right) \dy,
\end{equation}
where $c_1 = S$, $c_k = 1 +S $, for $2 \leq k \leq m-1$, and $c_m = 1$.
The STCDO spread $S^{*}_t$ at time $t\in[0,\delta]$ is equal to
 $$
 S^*_t= \frac{\sum_{k=1}^{m-1}\int_{x_1}^{x_2}   P(t, T_{k+1}) (F(t, T_k, y) - F(t, T_{k+1}, y)) \dy}
 {\sum_{k=1}^{m-1}\int_{x_1}^{x_2}  P(t, T_{k})F(t, T_{k}, y)\dy }.
 $$
\end{corollary}

\begin{proof}
Follows by inserting \eqref{eq:STCDO-cond-indep} into \eqref{eq:STCDO} and \eqref{eq:spread}.
\end{proof}

\begin{remark}
Corollary \ref{cor:STCDO-cond-indep} shows that under conditional independence of the default-free bond prices and the loss process, the STCDO spreads are given in terms of the initial term structure of the default-free bond prices and the $(T_k, x)$-forward prices.
This allows one to extract $(T_k,x)$-forward prices from market data.
\end{remark}

\subsection{Options on a STCDO}

 Consider a STCDO as defined in the previous subsection. Let us study an option which gives the right to enter into such a  contract at
 time $T_1$ at a pre-specified spread $S$. This is equivalent to a European call  on the STCDO with payoff
$$
\big( \pi^{STCDO} (T_1, S) \big)^+
$$
at $T_1$.  Assume that $(\mathbf{A5})$ holds. The value of the European call at time $t\in [0, \delta]$ is given by the
 expectation under the forward measure $\Q_{T_1}$:
\begin{align*}
\pi^{call} (t, S) & = P(t, T_1) \E_{\Q_{T_1}} \Big( \big( \pi^{STCDO} (T_1, S) \big)^+ \,|\,\cG_t \Big) \\
& = P(t, T_1)\E_{\Q_{T_1}} \left( \bigg(\int_{x_{1}}^{x_{2}} \Big( S \sum_{k=1}^{m-1} P(T_1, T_{k},y) - \sum_{k=1}^{m-1} e(T_1, T_{k+1}, y)\Big) \dy \bigg)^+ \,\Big|\,\cG_t \right),
\end{align*}
since by \eqref{eq:STCDO},
$$
\pi^{STCDO} (T_1, S) = \int_{x_{1}}^{x_{2}} \Big( S \sum_{k=1}^{m-1} P(T_1, T_{k},y) - \sum_{k=1}^{m-1} e(T_1, T_{k+1}, y)\Big) \dy.
$$
Assuming for simplicity that $P(t, T_k)=1$, for all $T_k$ and $t\leq T_k$, which implies the conditional independence which is assumed in Corollaries \ref{cor:cond-indep} and \ref{cor:STCDO-cond-indep}, we obtain
\begin{align*}
\pi^{call} (t, S)
& = \E_{\Q^*} \left( \bigg(\int_{x_{1}}^{x_{2}} \sum_{k=1}^{m} d_k F(T_1, T_k, y) \dy \bigg)^+ \,|\,\cG_t\right),
\end{align*}
where $d_1 = S-1$, $d_k = S$, for $2\leq k \leq m-1$, and $d_m=1$, which follows from \eqref{eq:STCDO-price-cond-indep}. Note that the measure $\Q_{T_1}$ coincides with the terminal forward measure $\Q^*= \Q_{T_n}$, cf. Remark \ref{rem:constant-risk-free}. Recall that
\begin{align*}
F(T_1, T_k, y) & =  F(0, T_k, y)\exp \Bigg( \sum_{i=1}^{k-1} \int_0^{T_1} a(t,T_i,y\big) \dt
 \\
&+   \sum_{i=1}^{k-1}\int_0^{T_1} b(t,T_i,y)  dX_t  +  \sum_{i=1}^{k-1} \int_0^{T_1} \int_{\cI}    c(t,T_i,y;z) \mu^L(dt,dz)\Bigg) \ifL{T_1}{y},
\end{align*}
for $k\geq 2$ and $F(T_1, T_1, y) = F(0, T_1, y) \ifL{T_1}{y}$.
We further assume $F(0, T_i, y)$, $a(t,T_i,y)$, $b(t, T_i, y)$ and $c(t, T_i, y; z)$ are constant in $y$ between $x_1$ and $x_2$. For simplicity we denote $a(t,T_i,y) = a(t,T_i,x_1)$ by $a(t,T_i)$ and similarly for the other quantities.
Then we have
\begin{align*}
\int_{x_{1}}^{x_{2}} \sum_{k=1}^{m} d_k F(T_1, T_k, y) \dy & = \sum_{k=1}^{m} d_k F(0, T_k)
\exp \Bigg(  \sum_{i=1}^{k-1} \int_0^{T_1} a(t,T_i) \dt
+  \sum_{i=1}^{k-1} \int_0^{T_1} b(t,T_i)  dX_t  \\
&+  \sum_{i=1}^{k-1} \int_0^{T_1} \int_{\cI}   c(t,T_i;z)  \mu^L(dt,dz)\Bigg) \int_{x_{1}}^{x_{2}}  \ifL{T_1}{y} \dy \\
& = f(L_{T_1})\sum_{k=1}^{m} d_k F(0, T_k)
\exp \Bigg(  \sum_{i=1}^{k-1} \int_0^{T_1} a(t,T_i) \dt  \\
& +   \sum_{i=1}^{k-1} \int_0^{T_1}  b(t,T_i)  dX_t
+   \sum_{i=1}^{k-1}\int_0^{T_1} \int_{\cI}  c(t,T_i;z)  \mu^L(dt,dz)\Bigg),
\end{align*}
for $f$ defined in \eqref{eq:f}. Note that $f:\cI \to \cI$, and so $f(L_{T_1})\ge 0$.
Thus, the value of the option at time $t$ is given by
\begin{align*}
\pi^{call} (t, S)
& =  \E_{\Q^*} \left( f(L_{T_1}) \Bigg(\tilde{d_1} + \sum_{k=2}^{m} \tilde{d_k}
\exp \bigg(  \sum_{i=1}^{k-1} \int_0^{T_1} a(t,T_i) \dt    \right. \\
& \left. +   \sum_{i=1}^{k-1} \int_0^{T_1}  b(t,T_i) dX_t
+  \sum_{i=1}^{k-1} \int_0^{T_1} \int_{\cI}  c(t,T_i;z)  \mu^L(dt,dz)\bigg) \Bigg)^+ \,\Big|\,\cG_t\right),
\end{align*}
where $\tilde{d_k}= d_k F(0, T_k)$, for $1 \leq k \leq m$.
 Assume now that $L$ and $X$ are conditionally independent given $\cG_t$. Therefore, if $c=0$ and $a(\cdot, T_i)$ and $b(\cdot, T_i)$ are conditionally independent of $L$ given $\cG_t$ for all $T_i$, this expression simplifies further to
\begin{align*}
\pi^{call} (t, S) & =  \E_{\Q^*} \left( f(L_{T_1}) |\cG_t\right) \E_{\Q^*} \left(
\Bigg(\tilde{d_1}  + \sum_{k=2}^{m} \tilde{d_k} \exp \bigg(  \sum_{i=1}^{k-1}  \int_0^{T_1} a(t,T_i) \dt \right.\\
& \left.\qquad +   \sum_{i=1}^{k-1} \int_0^{T_1} b(t,T_i)dX_t \bigg) \Bigg)^+ \,\Big|\,\cG_t\right),
\end{align*}
where
\begin{align*}
\E_{\Q^*} \left( f(L_{T_1}) |\cG_t\right) & = \E_{\Q^*} \left((x_2-x_1) \ifL{T_1}{x_1} + (x_2 - L_{T_1}) \indik_{\{x_1 < L_{T_1} \leq x_2\}} |\cG_t\right) \\
& = x_2 \Q^* \left( L_{T_1} \leq x_2 |\cG_t\right) - x_1 \Q^* \left( L_{T_1} \leq x_1 |\cG_t\right) \\
&- \E_{\Q^*} \left( L_{T_1} \indik_{\{x_1 < L_{T_1} \leq x_2\}} |\cG_t\right).
\end{align*}

As far as the second factor in $\pi^{call} (t, S)$ is concerned, it is similar to the expressions that appear in valuation formulas for swaptions in term structure models without defaults. It can be computed using Fourier transform techniques under appropriate technical assumptions; cf. \citeN{EberleinKluge06a} and \citeN{KellerResselPapapantoleonTeichmann09}. In particular, we refer to
\citeN{EberleinGlauPapapantoleon10} and \citeN{Eberlein12} for Fourier transform methods in a general semimartingale setting. For the affine specification given in Section \ref{sec:affine}, this approach may be simplified further.

\section{Calibration}
\label{sec:calibration}

In this section we give a calibration exercise with a two-factor affine diffusion
which on one side shows the flexibility of our framework
in a simple  specification and further illustrates the implementation of the model.
For the calibration, we use the affine model from Section \ref{sec:affine} and
implement an extended Kalman filter as suggested in \citeN{EksiFilipovic12}.
In contrast to typical calibration approaches we do not
only fit to data of single days but to the data of a period of two and a half years, namely from February  2008 to August 2010.
 The model is able to provide a surprisingly good fit across the different tranches and
maturities as we shall illustrate.

\subsection{The dataset}
The  calibration is performed on data from the iTraxx Europe index, more specifically  it consists of implied \emph{zero-coupon spreads} of the iTraxx Europe\footnote{We thank Dr. Peter Schaller for providing us the data.}.
In the market there are STCDOs on the iTraxx Europe with detachment points
 $\{x_1,\dots,x_J\}=\{0, 0.03, 0.06, 0.09, 0.12, 0.22, 1\}$.
The zero-coupon spreads are the quoted spreads of the STCDOs, %above the risk-free interest,
in our notation
given by
\begin{align}\label{def:R}
R(t,\tau ,j) &:= -\frac{1}{\tau}\log\bigg(\frac{1}{x_{j+1}-x_j}\int_{x_j}^{x_{j+1}}F(t,t+\tau,x)dx\bigg),
\end{align}
where $\tau$ denotes  time to maturity. In the data we have $\tau\in\{3,5,7,10\}$.
In the model we consider later the case where $F(t,T,x)$ is constant  in the intervals $[x_j,x_{j+1})$ and then
$$ - \tau \cdot R(t,\tau ,j) = \log P(t,t+\tau,x_j) - \log P(t,t+\tau)$$
as $F(t,T,x)=P(t,T,x) \, P(t,T)^{-1}$ by definition. Therefore, the rate $R$ indeed refers to a spread above the  risk-free rate.

The realized index spreads are shown in Figure \ref{fig:indexspread}. With the beginning of the credit crisis, volatility,  as well as the credit spreads, jumped
to very high levels stabilizing thereafter. In the first quarter of 2010  a new increase due to the European debt crises can easily be spotted. Figure \ref{fig:dblpic} shows the evolution of the tranche spreads for different maturities and tranches. The spread curves follow a similar pattern. Consequently, it is plausible to capture the dynamics with a low number of factors.
It is important to mention that in the observation period defaults did not occur in the underlying pool.

\begin{figure}
\includegraphics[width=0.6\textwidth]{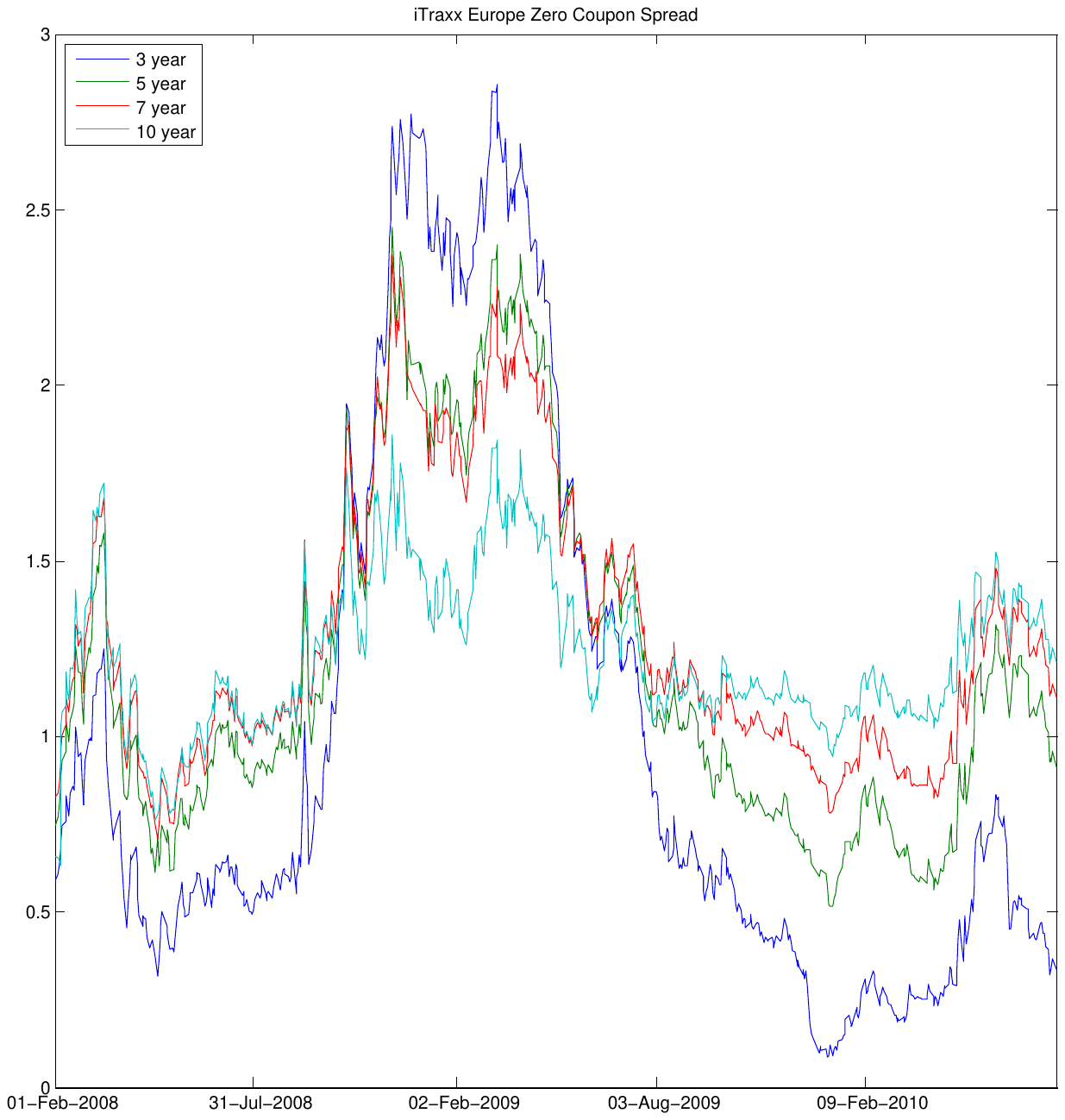}\\[-1.3cm]
\caption{The {iTraxx Europe} zero-coupon index spread for the period February 2008 to August 2010.
The different graphs refer to the time to maturity of 3,5,7 and 10 years.}
\label{fig:indexspread}
\end{figure}

\begin{figure}
\includegraphics[width=0.6\textwidth,height=8cm]{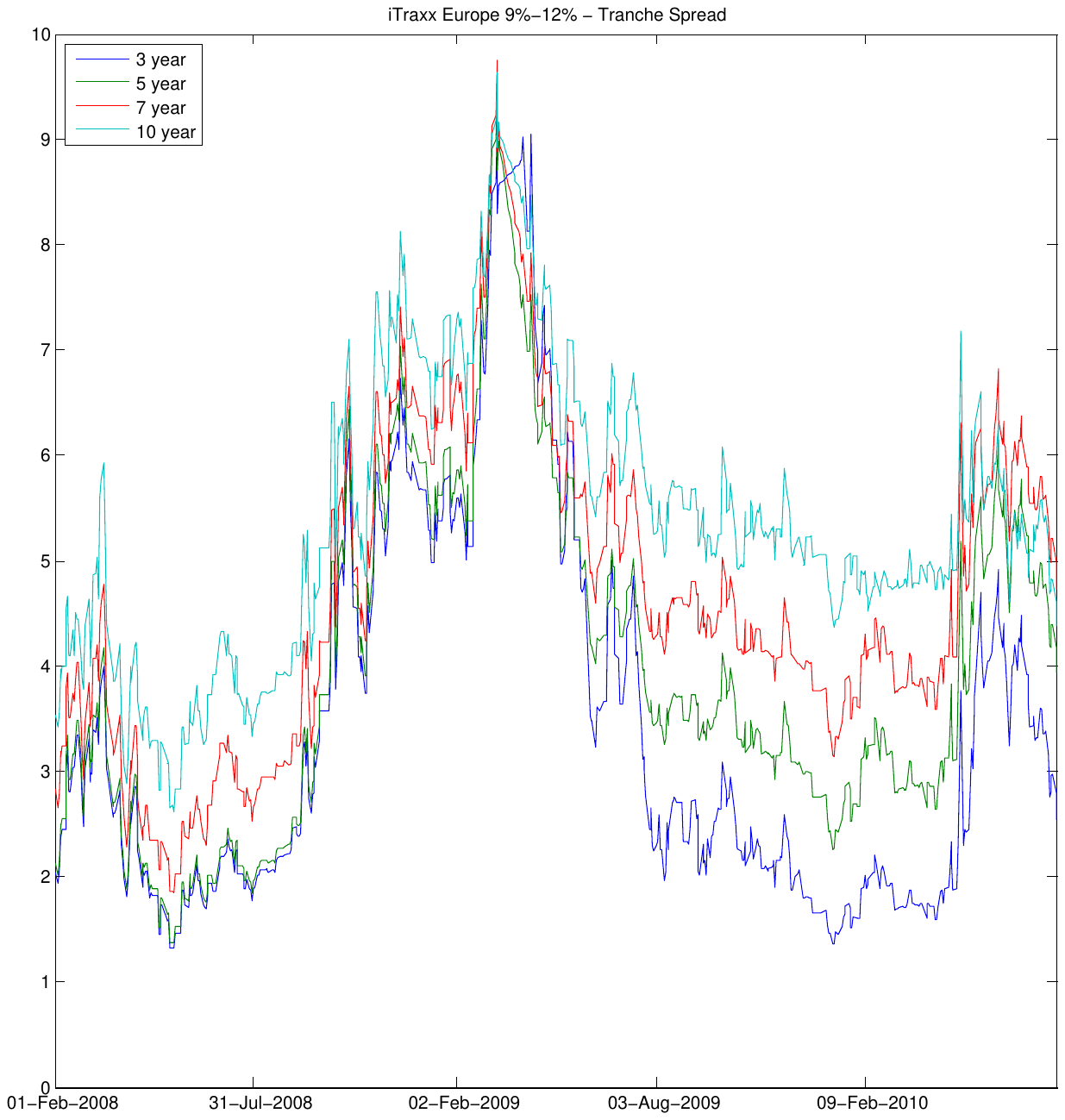} \\[-1.5cm]
\includegraphics[width=0.6\textwidth,height=8cm]{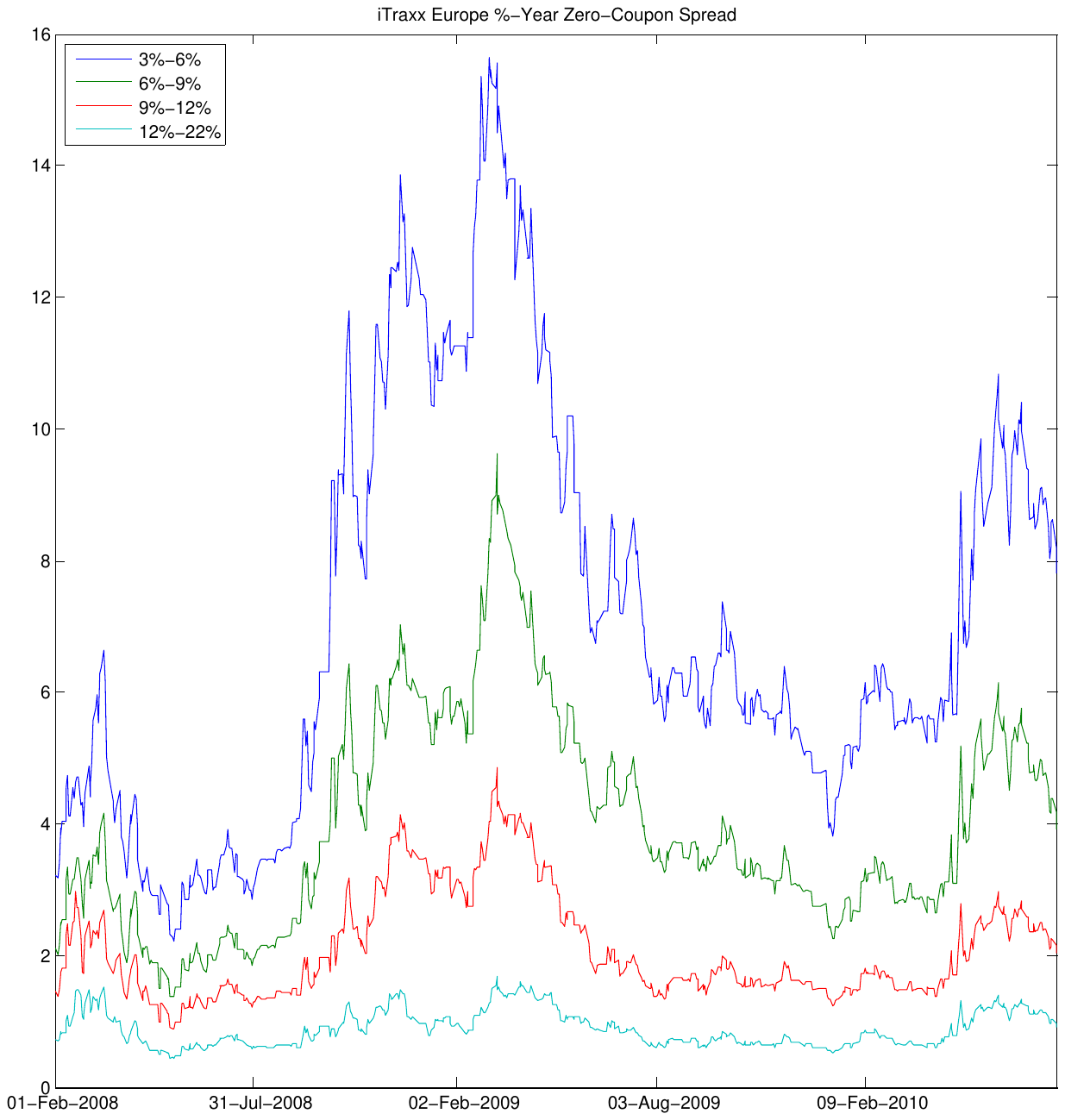}\\[-1.1cm]
\caption{
The upper graph shows the iTraxx Europe 9\%-12\% tranche spread from February 2008 to August 2010 for different maturities. The lower graph illustrates the iTraxx Europe tranche spreads from February 2008 to August 2010 for a fixed maturity of five years.}
\label{fig:dblpic}
\end{figure}

\subsection{Model specification}
Our aim is to calibrate a simple two-factor affine diffusion model to the \emph{whole} data set using Kalman filtering.
To this end, we specify the model under the \emph{physical} probability measure $\P$. Prices of traded products
are computed under the \emph{risk-neutral} measure $\Q$ which we obtain by a change of
measure where  the affine structure is kept.

A principal component analysis reveals that two factors already explain 88.30\% of the realized variance; see \citeN{EksiFilipovic12}. We therefore consider a two-dimensional affine process $Z=(Z^1,Z^2)^\top \in\R^+\times\R^+=: \cZ$ satisfying
\begin{align*}
dZ^1_t &= \kappa^1(Z^2_t-Z^1_t)dt + \sigma^1\sqrt{Z^1_t}dW^1_t\\
dZ^2_t &= \kappa^2(\theta^2-Z^2_t)dt + \sigma^2\sqrt{Z^2_t}dW^2_t,
\end{align*}
and $Z_0=(z_1,z_2) \in \cZ$.
Here $\kappa^1,\kappa^2, \theta^2, \sigma^1,\sigma^2$ are positive constants and $W^1$ and $W^2$ are
independent standard Brownian motions. The factor $Z^2$ is the stochastic mean reversion level of $Z^1$.

For the measure change we specify the market prices of risk  by
$$\lambda_t^i = \frac{\lambda^i\sqrt{Z_t^i}}{\sigma^i}, \qquad i=1,2 $$
with constants $\lambda^1,\lambda^2 \in \R$.
Using Girsanov's theorem, we change to an equivalent measure $\Q$ where
$\tilde{W}^i_t = W^i_t + \int_0^t\lambda_s^i ds$, $i=1,2$ are independent standard Brownian motions.
Then, under  $\mathbb{Q}$, $Z$ is again affine and satisfies the following
dynamics; see  \citeN{CheriditoFilipovicKimmel2010}:
\begin{align*}
dZ^1_t &= (\kappa^1+\lambda^1)\left(\frac{\kappa^1}{\kappa^1+\lambda^1}Z^2_t-Z^1_t\right)dt + \sigma^1\sqrt{Z^1_t}d\tilde{W}^1_t,\\
dZ^2_t &= (\kappa^2+\lambda^2)\left(\frac{\kappa^2}{\kappa^2+\lambda^2}\theta^2-Z^2_t\right)dt + \sigma^2\sqrt{Z^2_t}d\tilde{W}^2_t.
\end{align*}
Hence, $Z$ is an affine process under $\Q$ and we may apply the results from Section \ref{sec:affine}.

For a complete specification of the model we  need to specify the compensator of the loss process $L$
and the contagion parameter $c$.
According to our setup we assume that $m$ depends in an  affine way on $Z$ and we assume that it is only
driven by $Z^1$, i.e.
$$
m(t,l,z,dy)= m_0(t,l,dy) + m_1(t,l,dy) z_1.
$$
We choose the jump distribution from the beta family, more precisely
$$ m(t,l,z,dy)
= \frac{1}{B(a_1,b_1)}y^{a_1-1}(1-y)^{b_1-1}dy + \frac{z_1}{B(a_2,b_2)}y^{a_2-1}(1-y)^{b_2-1}dy,
$$
where all coefficients are positive.
Finally we specify the contagion parameter and assume that
\begin{equation}
c(t,T_k,x,L_{t-};y) = cy(T_k-t).
\end{equation}

We consider $H$  specified as in \eqref{eq:Haffine} together with  \eqref{eq:affineA} and \eqref{eq:affineB}
and it follows from Proposition \ref{prop:affine} that
this is an arbitrage-free model.

\subsection{The calibration procedure}
For the estimation of the (unobserved) variables $Z$ from the observed STCDO prices we use an extended
Kalman filter following \citeN{EksiFilipovic12}.
Furthermore we make the following two
assumptions: first, we assume that tranche spreads are piecewise constant between the detachment
points, that is
$$H(t,T_k,x)= H(t,T_k,x_{i+1}), \quad\text{ for } x\in[x_{i},x_{i+1}).$$
Second, we assume that  observed prices are given by model implied prices with additive noise.
More formally, we assume that at observation times $0=t_0,t_1,t_2,\dots$
\begin{align*}
R(t_k,\tau,j)
&= -\frac{1}{\tau}\log\bigg(\frac{1}{x_{j+1}-x_j}\int_{x_j}^{x_{j+1}}F(t_k,t+\tau,x)\, dx\bigg)
+ \varepsilon(k,\tau,j+1) \\
&=:\alpha(\tau,x_{j+1}) - \frac{1}{\tau}\beta(\tau,x_{j+1})Z_{t_k} - c L_{t_k} + \varepsilon(k,\tau,j+1).
\end{align*}
Note that with $H$ also $F$ is affine. Moreover, as $A$ and $B$ are piecewise constant, the terms $\alpha$ and $\beta$ are straightforward to
compute, see \citeN{GehmlichGrbacSchmidt12} for detailed computations.
The measurement error consists of independent and normally distributed random variables,
where the variance of the measurement errors may differ across the observed tranches:
\,$\varepsilon(k,\tau,j+1)\sim \cN(0,\sigma_{j+1})$.

We approximate {the conditional distribution of $Z_{t_k}$ given
$Z_{t_{k-1}}$ }by a normal distribution
where the first and the second moments are matched. This is in line with a quasi-maximum-likelihood approach
and simplifies the computations considerably. The moments of the affine diffusion $Z$ can be
computed using the Kolmogorov backward equation; see  Proposition 3.1 in \citeN{EksiFilipovic12}.
This enables us to apply the extended Kalman filter algorithm to obtain a calibration to the
full data set.
The  details of this approach and the extension to more factors can be found in \citeN{GehmlichGrbacSchmidt12}.

\begin{remark}
As an alternative to the filtering approach which we favor here one could also use nonlinear least squares  to fit the model to data.
Such an approach is pursued in  \citeN{LongstaffRajan08}, notably on a quite different model.
They fit the unknown parameter vector $\theta$, as well as the unobserved factor process $Z$, to the data by minimizing
the sum of squared distances between the observed prices and the model prices computed  with parameter $\theta$ and  the factor process $Z_1,\dots,Z_T$ taking values $z_1,\dots,z_T$.
Applying this procedure to market data of the CDX NA IG for the period from October 2003 to October 2005 they fit a three-factor model.
A comparison to the filtering approach reveals on one side, that nonlinear least squares give only access to the parameters under the risk-neutral measure. On the other side,
%which can be formulated as follows:
% denote by $\hat H_i(t,\btheta,z)$, $i=1,\dots,I$ the various tranche spreads implied by the model with parameter set $\btheta$ and given that the factor process $Z_t$ takes the value $z$.
% The unkown parameter vektor $\btheta$ and the unobserved factor process $Z$ are estimated form observed tranche spread data $H(t_1),\dots,H(t_N)$ by solving the minimization problem
% $$ \min_{\btheta,z_{t_1},\dots,z_{t_N}} \sum_{l=1}^N\sum_{i=1}^I (H_i(t_j) - \hat H_i(t_j,\btheta,z_j))^2. $$
%They obtain a comparable fit to market data of the CDX NA IG for the period from October 2003 to October 2005 with a three-factor model.
with the filtering approach one gets additional regularity on the estimated factor process in comparison to nonlinear least squares.  In this regard, it is surprising that the model considered here is able to provide an excellent fit to a longer and more turbulent time series with two factors only. For details we refer to the calibration results in the following section.
\end{remark}

\subsection{Calibration results}

The extended Kalman filter allows a calibration to the full dataset from February 2008 to August 2010. On one side, the Kalman filter provides an estimation of  the hidden state process $Z$  and
on the other side, maximizing the quasi-likelihood function given the estimated values of $Z$, gives
the estimator of the parameter vector. Table \ref{table1} shows the estimated values.

\begin{table}[h]

\begin{center}
\tiny{
\begin{tabular}{@{} c*{12}{ c} @{}}
\toprule
 $\lambda^1$ & $\lambda^2$ & $\kappa^1$ & $\kappa^2$ & $\theta^2$ & $\sigma^1$ & $\sigma^2$ & $c$ & $a_1$ & $b_1$ & $a_2$ & $b_2$\\
\midrule
-0.0780 & -2.5472 & 1.5722 & 1.8569 &  0.4720 &  0.7305 & 0.1739 &  -0.0571&  0.6797 & 5.1597 &0.2492 & 22.26 \\
\bottomrule
\end{tabular}}

\vspace{2mm}

\end{center}
\caption{Estimated parameter values. \label{table1}}
\end{table}

It turns out, that the jump distribution in $m_1$
is quite close to an  exponential distribution,
as $a_2$ is small. However, $a_1$ contributes significantly to the  fit of the model. The contagion parameter $c$ is negative, as expected --
an occurring loss, i.e.~an upward jump in the loss process leads to a downward jump in the $(T,x)$-bond prices
by a downward jump in $H$.

Based upon the estimated parameter values and the filtered factor process we regenerate the data.
In Figures \ref{cal1} and \ref{cal2} we plot estimated vs. observed values. For brevity, the longest maturity which shows a similar behavior is left aside. The graphs can be used for the diagnosis
of the model fit. It is remarkable that the two-factor model is able to provide an excellent fit
across all tranches and over the whole data period. This underlines the stability of the approach which
leads to improved hedging performance, as shown in \citeN{EksiFilipovic12}. They obtain a similar fit
with a two-factor affine model when incorporating additionally a catastrophic component. As pointed out,
a two-factor model with zero catastrophic component is not able to provide a good fit to the super-senior tranche.
In our approach, the additional freedom obtained by considering a discrete tenor structure allows to incorporate a contagion term which improves the fit substantially. Compare in particular Figure \ref{cal2}.

\begin{figure}
\begin{tabular}{@{} c*{2}{c} @{}}
\hspace{-2cm}
	\includegraphics[width=0.6\textwidth,height=14cm]{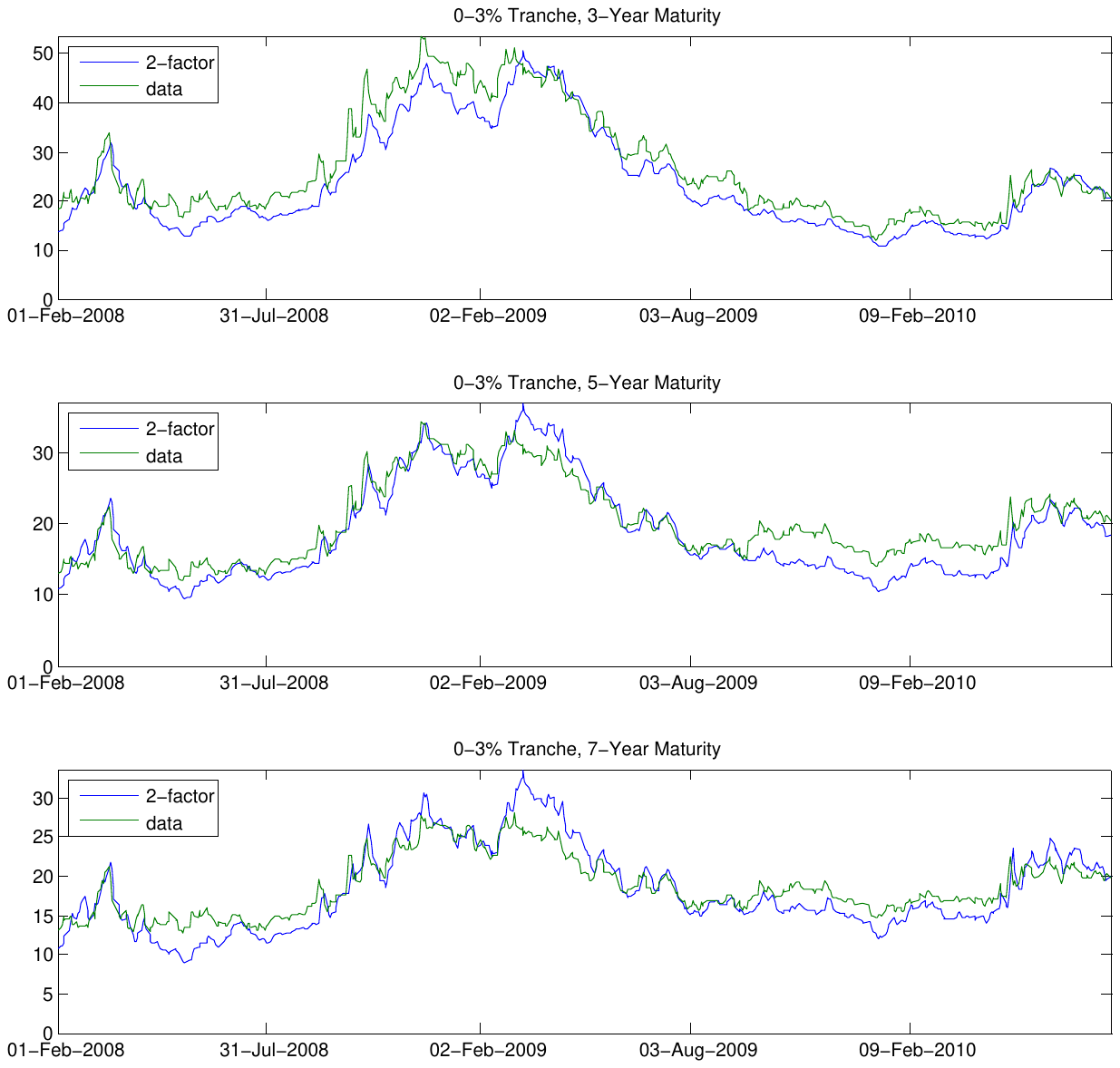} &
	\includegraphics[width=0.6\textwidth,height=14cm]{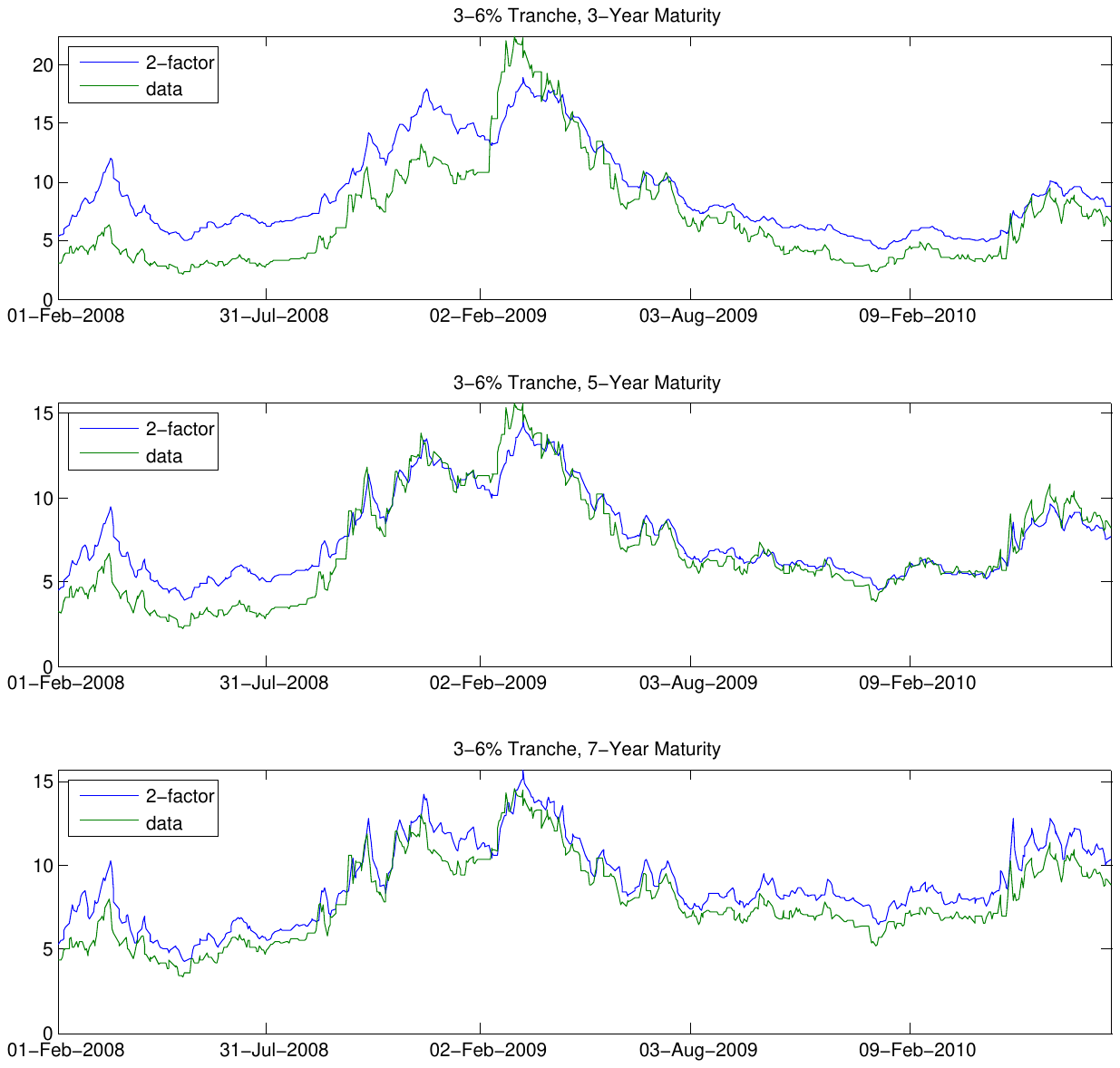}\\[-2.3cm]
0\%-3\% Tranche   & 3\%-6\% Tranche    \\[-1.5cm]
\hspace{-2cm}
	\includegraphics[width=0.6\textwidth,height=14cm]{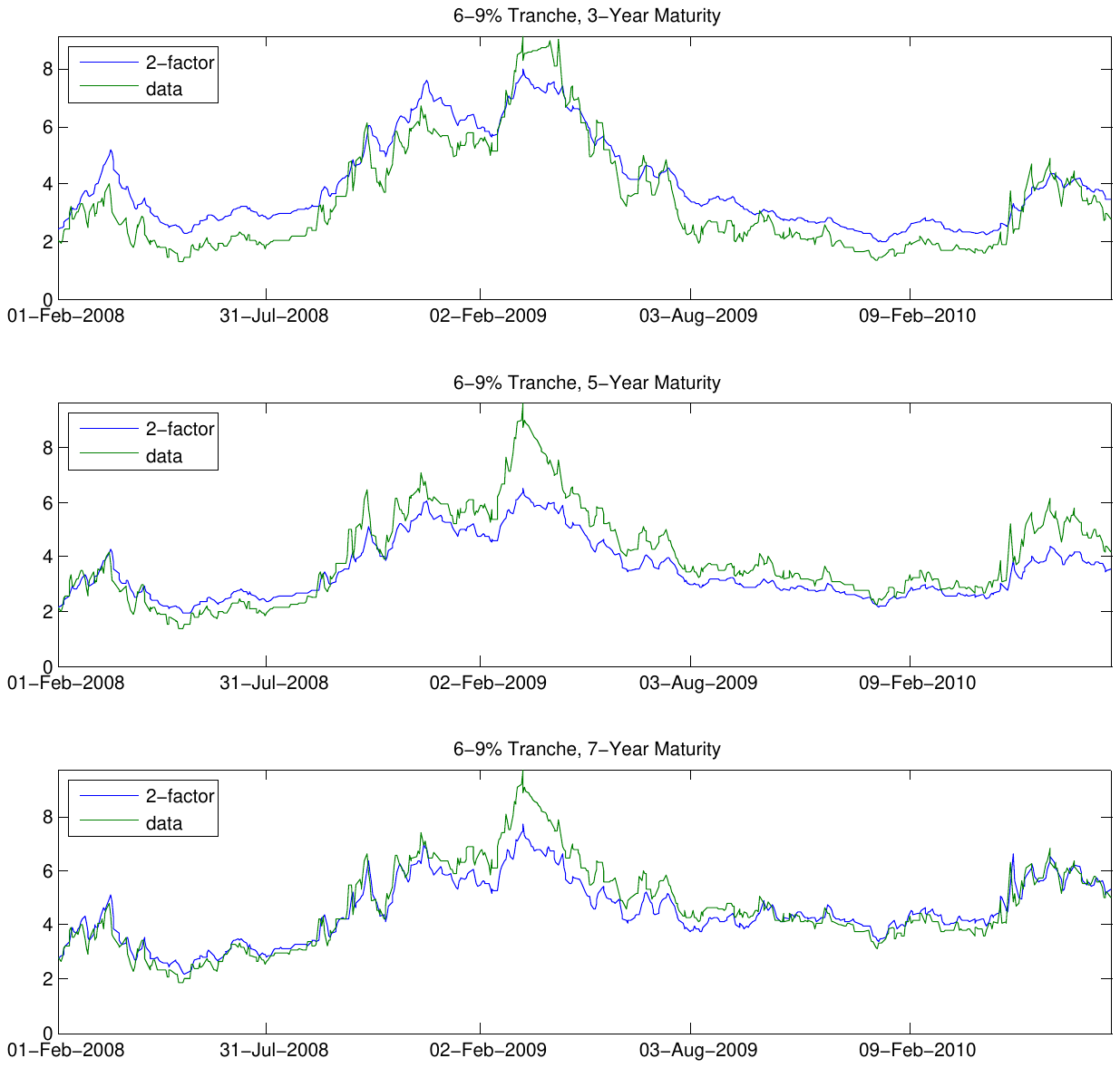} &
	\includegraphics[width=0.6\textwidth,height=14cm]{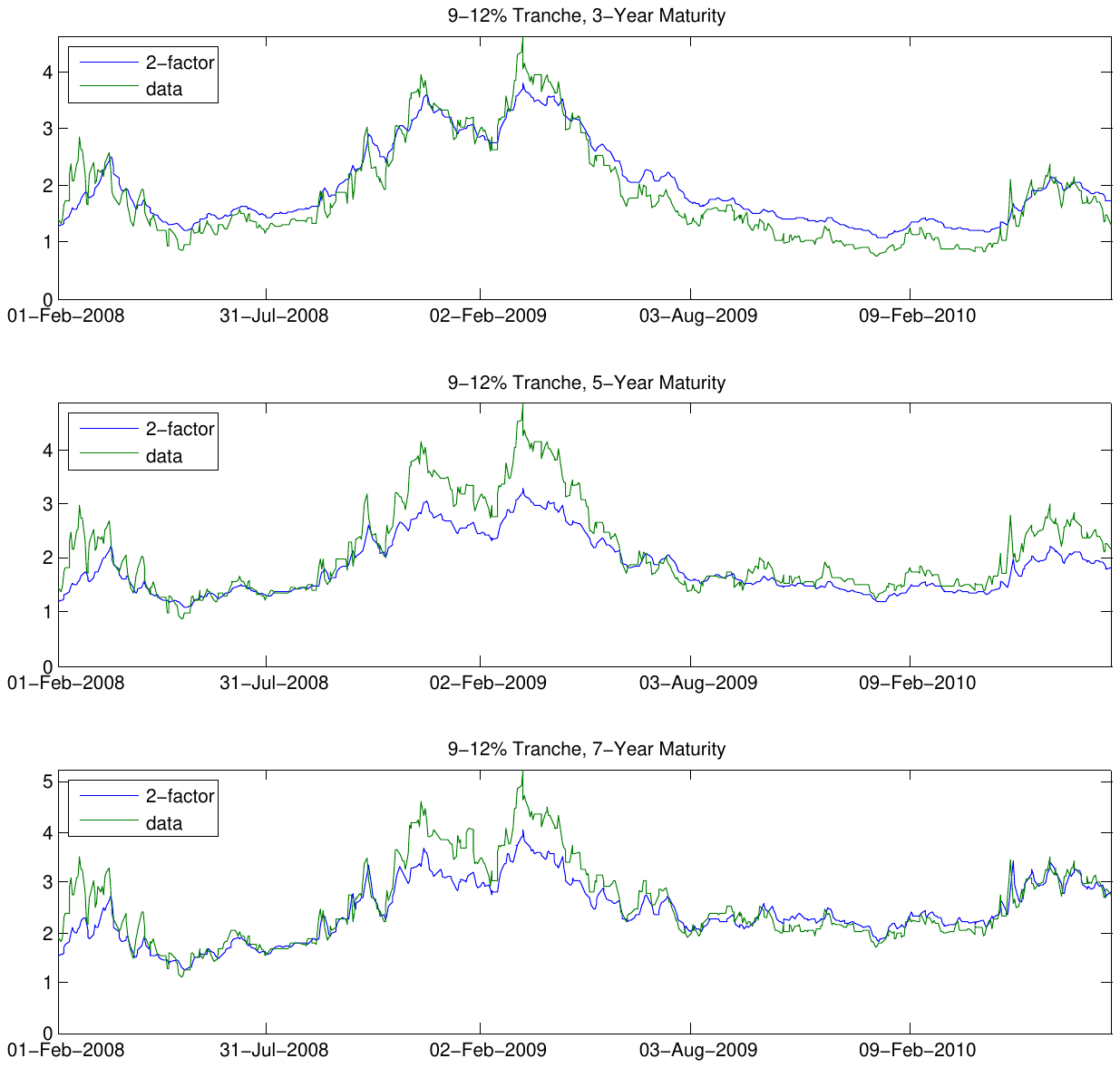}\\[-2cm]
6\%-9\% Tranche    & 9\%-12\% Tranche     \\
\end{tabular}
\caption{Estimated and realized data - part 1.\label{cal1}}
\end{figure}

\clearpage

\begin{figure}
\begin{tabular}{@{} c*{2}{c} @{}}
\hspace{-2cm}
	\includegraphics[width=0.6\textwidth,height=14cm]{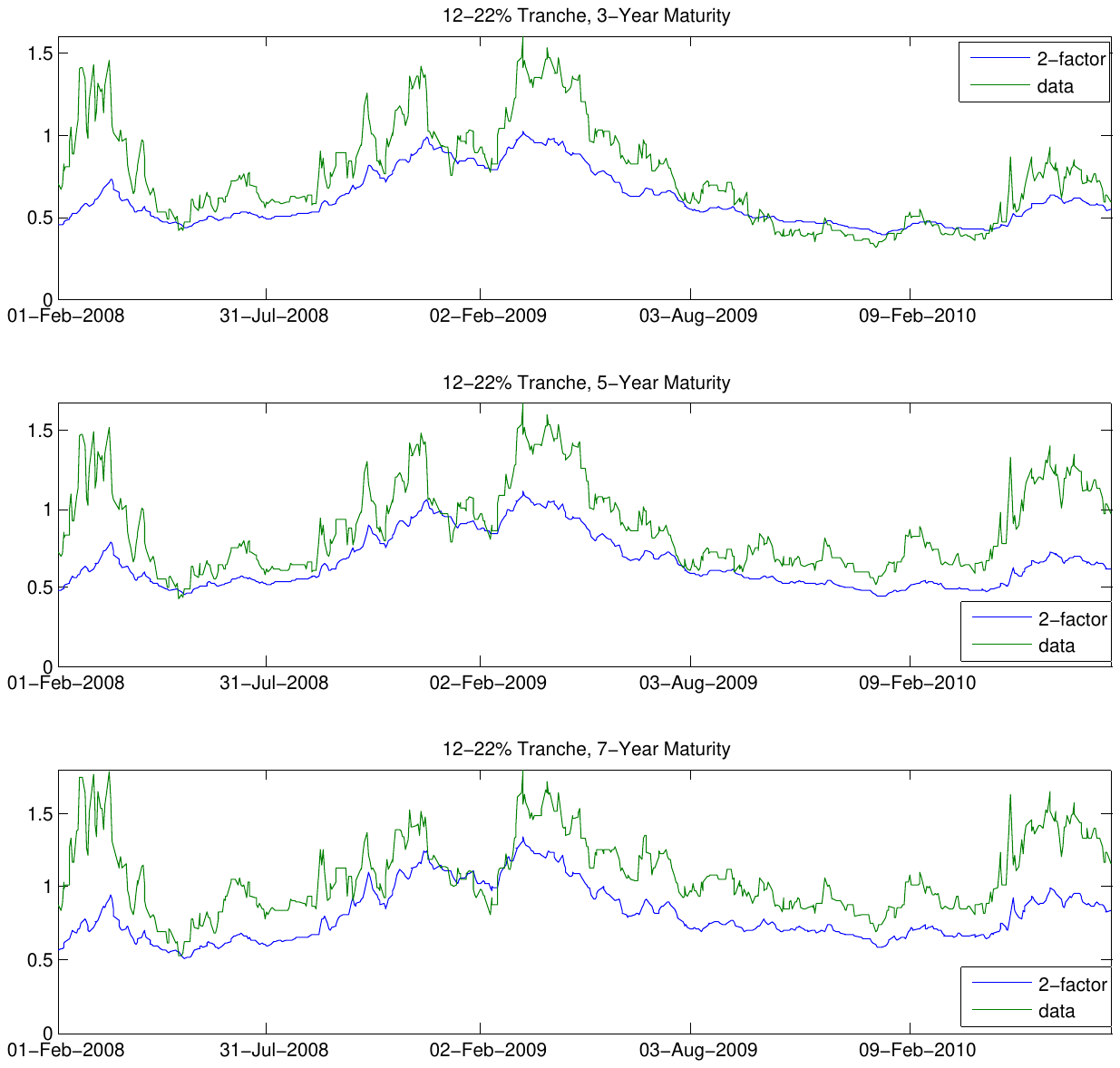} &
	\includegraphics[width=0.6\textwidth,height=14cm]{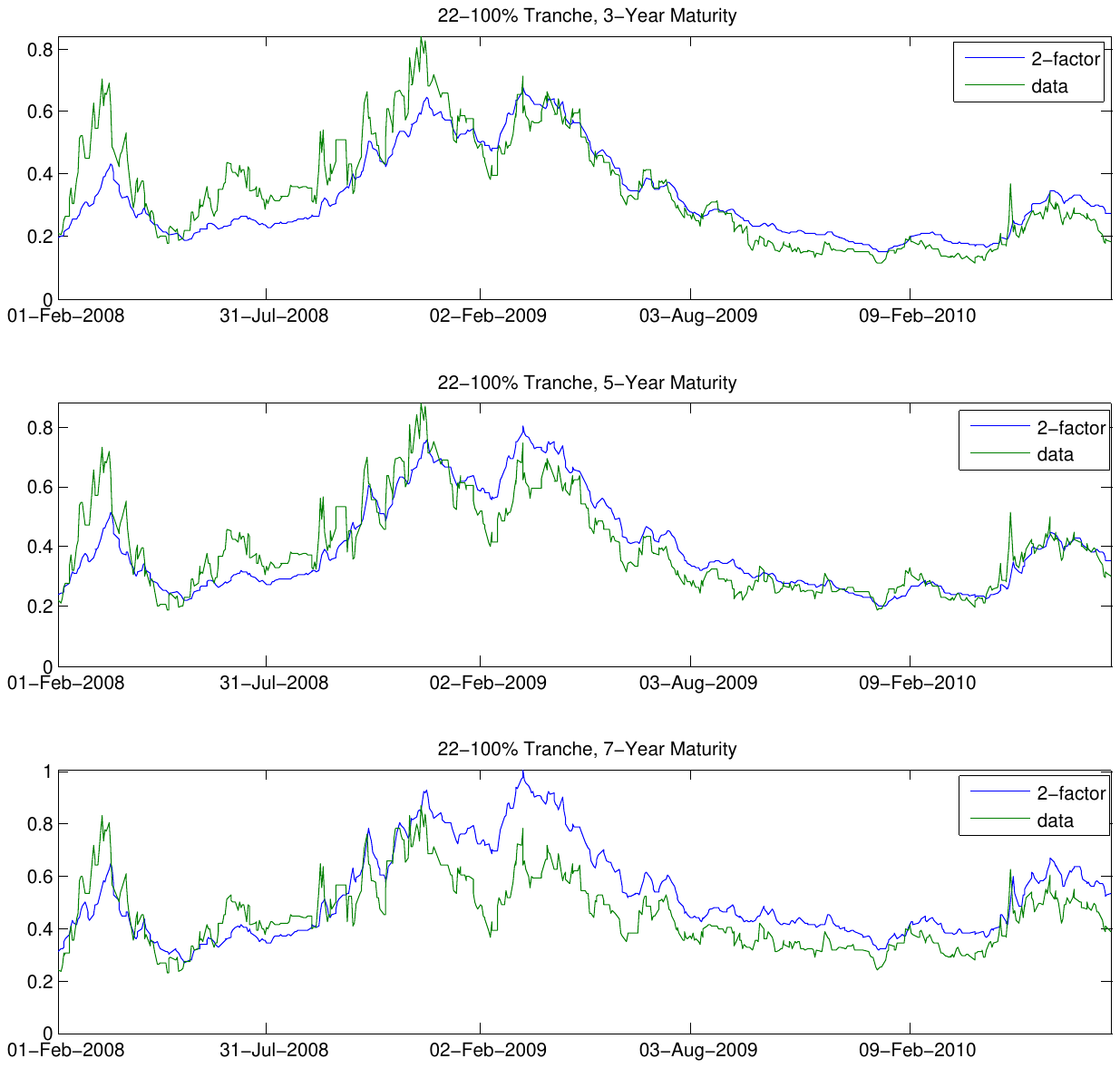}\\[-2.3cm]
12\%-22\% Tranche   & 22\%-100\% Tranche
\end{tabular}
\caption{Estimated and realized data - part 2.\label{cal2}}
\end{figure}
\appendix

%\bibliographystyle{chicago}      % mathematics and physical sciences
%\bibliography{biblio}   % name your BibTeX data base

% Update bibliography. We have once Libor in the bibliograhpy. Is that correct ?

\end{document}